\documentclass[prd,aps,eqsecnum,floatfix,nofootinbib,preprint,tightenlines]{revtex4}
\usepackage{dcolumn}
\usepackage{latexsym}
\usepackage{graphicx}
\usepackage{bm}

\def\b{\beta}

\def\d{\delta}
\def\g{\gamma}

\def\j{\psi}

\def\l{\lambda}
\def\m{\mu}

\def\p{\pi}                     
\def\th{\theta}                  
\def\r{\rho}                    
\def\s{\sigma}                  
\def\t{\tau}

\def\G{\Gamma}
\def\J{\Psi}

\def\cc{{\cal C}}
\def\cd{{\cal D}}

\def\cg{{\cal G}}

\def\cj{{\cal J}}

\def\co{{\cal O}}
\def\cp{{\cal P}}

\def\car{{\cal R}}

\def\cu{{\cal U}}

\def\dg{^\dagger}                                     
\def\half{{1\over 2}}
\def\svev#1{\left\langle #1\right\rangle}       
\def\vev#1{\Big\langle #1 \Big\rangle}           
\def\sbraket#1#2{\left\langle#1|#2\right\rangle} 
\def\eg{\mbox{\it e.g.} }

\def\bibi{\bibitem}

\def\bj{\overline\psi}
\def\Tr{{\rm Tr\,}}
\def\db{\partial^*}

\def\tG{\tilde\G}
\def\tR{\tilde{\cal R}}

\def\mres{m_{\rm res}}

\def\tH{\tilde{H}}

\def\Gimp{\Gamma_{\rm imp}}
\def\dov{D_{\rm ov}}
\def\lov{l_{\rm ov}}

\begin{document}

\title{Localization properties of lattice
fermions with\\ plaquette and improved gauge actions}
\author{Maarten Golterman}%
 \affiliation{Department of Physics and Astronomy,
San Francisco State University, San Francisco, CA 94132, USA}

\author{Yigal Shamir}
\author{Benjamin Svetitsky}
 \affiliation{School of Physics and Astronomy, Raymond and Beverly
Sackler Faculty of Exact Sciences, Tel~Aviv University, 69978
Tel~Aviv, Israel}

\begin{abstract}
We determine the location $\lambda_c$ of the mobility edge
in the spectrum
of the hermitian Wilson operator in pure-gauge ensembles
with plaquette, Iwasaki, and DBW2 gauge actions.
The results allow mapping a portion of the (quenched) Aoki phase diagram.
We use Green function techniques to study the localized and extended
modes.
Where $\lambda_c>0$ we characterize
the localized modes in terms of an average support length and an
average localization length, the latter determined from
the asymptotic decay rate of the mode density.
We argue that, since the overlap operator is commonly
constructed from the Wilson operator, its range
is set by the value of $\lambda_c^{-1}$ for the Wilson operator.
It follows from our numerical results
that overlap simulations carried out with a cutoff of 1~GeV, even with improved gauge actions,
could be afflicted by unphysical degrees of freedom as light as 250~MeV.
\end{abstract}


\maketitle

\section{Introduction}
\label{Int}

Domain-wall and overlap fermions
reconcile chiral symmetry with the lattice, allowing for exact chiral
symmetry at finite lattice spacing in the euclidean path-integral
\mbox{formulation~\cite{dbk,dwf,oovlp,ovlp,GWL}}.
While chiral symmetry can be achieved for a range of non-zero
bare coupling $g_0$, problems arise if the bare coupling is too large.
For domain-wall fermions, chiral symmetry cannot be maintained
in the strong-coupling limit \cite{BBS}.
For overlap fermions, the built-in (modified) chiral symmetry is exact,
but at strong coupling one loses either locality \cite{HJL,lclz} or control
over the number of species \cite{ovlp,hopp}.

In any numerical simulation
it is important to stay away from the dangerous regions of the phase diagram.
The lattice Dirac operators of domain-wall fermions (DWF)
and overlap fermions are both based on a Wilson operator with a negative,
super-critical bare mass $m_0$.\footnote{
  The super-critical region is $-8<am_0<0$.
Outside of this region the Wilson operator cannot
have zero eigenvalues.}
  Locality and chirality in these formulations are controlled
by the spectral properties of this underlying Wilson operator.

\begin{figure}[thb]
\begin{center}
\includegraphics*[width=.7\columnwidth]{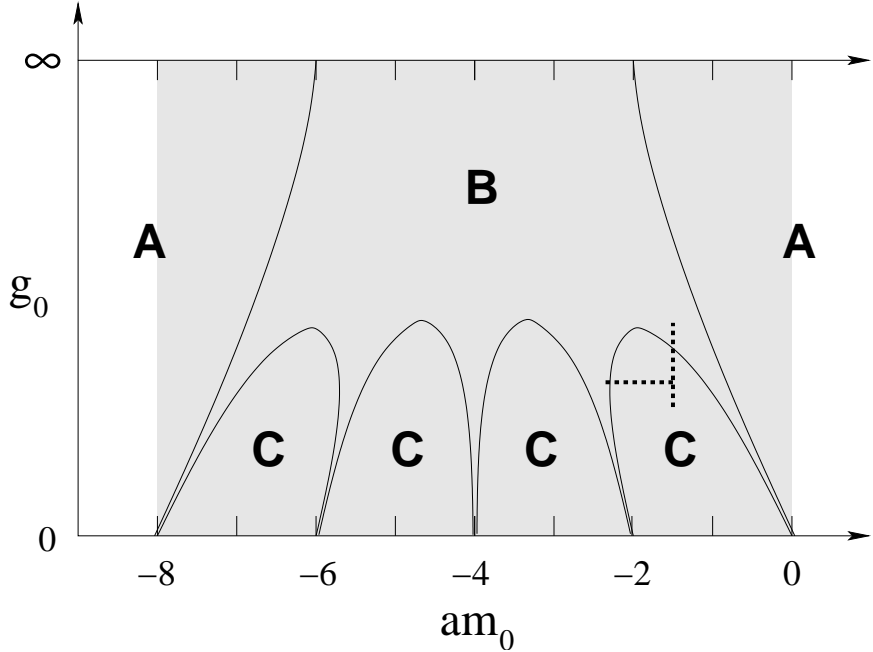}
\caption{A schematic representation of the Aoki phase diagram.
The lightly shaded area is the super-critical region.
Phase B is the (massless) Aoki phase, while phases A and C are massive.
Depending on the action in the theory,
each of the thin ``fingers'' where the Aoki
phase touches the line $g_0=0$ could alternatively be replaced by a line
of first-order phase transition \cite{ShSi}. Recent numerical results
support the latter scenario in the case of dynamical Wilson
fermions \cite{nofinger}.
The phase diagram is symmetric under the replacement
$am_0\to -(8+am_0)$, which can be undone by redefining the fields.
Dotted lines indicate where we performed our numerical analysis.
\label{aoki_crude}}
\end{center}
\end{figure}

The outstanding features of the super-critical Wilson operator are best
illustrated in a theory with
two dynamical flavors of Wilson fermions.
Here the absence of a spectral gap in part of the phase
diagram implies the existence of
propagating, light degrees of freedom.
Moreover, a non-zero spectral density for vanishing eigenvalue
signals spontaneous symmetry breaking,
as follows from the Banks--Casher relation \cite{BC}.
For Wilson fermions there is no chiral symmetry to be broken, so
the spontaneously broken symmetry is vectorial.
As discovered by Aoki \cite{aoki}, the pions become massless
if the bare Wilson-quark mass $m_0$ is lowered from positive
values towards a critical value
$m_0=m_c(g_0)<0$. For $m_0<m_c(g_0)$ the curvature of the effective potential
for pions becomes negative at the origin; a pionic condensate forms which breaks
spontaneously isospin and parity. This is the Aoki phase.
Inside the Aoki phase the condensing pion is massive,
while the other two pions are Goldstone bosons of the spontaneously
broken isospin generators. See Fig.~\ref{aoki_crude}
for a schematic representation of the phase diagram.

For DWF or overlap fermions one aims for a big gap, of order $1/a$,
in the spectrum of the Wilson operator.
Simply speaking, a bigger gap in the spectrum of the Wilson operator improves
both the chiral symmetry of DWF (at fixed, finite extent of the fifth dimension)
and the locality of overlap fermions. (As we will shortly see, in reality
any such gap can be mostly, but not completely, devoid of eigenvalues.)
Assuming that the phase diagram of the
underlying Wilson operator remains qualitatively the same as
in  Fig.~\ref{aoki_crude}, this requires being inside
one of the C phases. In practice, the rightmost C phase is used which,
for $g_0\to 0$, coincides with the interval $-2<am_0<0$.
In this interval, one lattice domain-wall (or overlap) field gives
rise to one quark in the continuum limit.

When one studies the spectral properties of the Wilson operator as a kernel for DWF or overlap fermions, one is in fact considering a {\em quenched\/} Wilson-fermion theory, because the Boltzmann weight is derived from a different fermion operator.
Usually, quenching means leaving out the fermion
determinant altogether, but we will also consider the more general sense that the fermion determinant is that of DWF or the overlap operator.
In this paper, we will study the spectrum of the Wilson operator for a variety
of pure-gauge theories, but we will take our results as indicative of
what might arise in a theory with dynamical
fermions.

The Goldstone theorem connects spontaneous symmetry breaking with the
appearance of massless poles in correlation functions.
Quenching the Wilson fermions, however, opens the door to a
new dynamical possibility.
Considering again a two-flavor theory,
let us suppose that all the correlation functions of the two (quenched) Wilson
flavors decay exponentially.
(As we will see, this is indeed the case well inside the C phases.)
In the quenched theory, this does {\em not\/} preclude a non-zero
pion condensate:
The isospin symmetry can be broken spontaneously {\em without\/} creating
a Goldstone boson.
It was shown in \cite{ms,lclz} how
this can be reconciled with the usual Ward-identity argument
for the existence of a Goldstone pole.

  There is now solid numerical and semi-analytical evidence \cite{SCRI,BNN}
for the existence of zero modes of the Wilson operator
throughout practically the entire super-critical region.
The Banks--Casher relation again
leads to a non-zero pion condensate.
In particular, the pion condensate is non-zero for parameter values
in the C phase that yield good-quality DWF and overlap-fermion simulations.
Despite the condensate, all the Wilson-fermion correlation functions are short ranged.

Ref.~\onlinecite{lclz} provides a theoretical explanation of this situation.
The low-lying eigenvectors of the hermitian Wilson operator may
be either extended or (exponentially) localized.
In the first case, the condensate must be accompanied by Goldstone bosons.
In the second case, there is an alternative mechanism for
saturating the relevant Ward identity, and all correlation functions
can be (and, in fact, are) short-ranged.
This gives the following physical picture
for the quenched Wilson-fermion phase diagram.%
\footnote{See also Ref.~\onlinecite{lt03}, in particular Fig.~2 therein.}
In the entire super-critical region there is no gap
in the spectrum of the Wilson operator, and the pion condensate is non-zero.
For eigenvalues $\l$ above a certain mobility edge $\l_c$,
the eigenvectors
of the Wilson operator are extended.
If $\l_c>0$, eigenvalues $|\l|<\l_c$
correspond to localized eigenmodes. In this case the pion condensate is non-zero, but
there are still no long-range correlations.
When $\l_c=0$, on the other hand, the condensate arises from extended eigenmodes,
and there are Goldstone pions.

The quenched Aoki phase is identified with
the region where Goldstone pions exist; that is, it is defined by
$\l_c=0$. With this definition, the quenched phase
diagram could be qualitatively similar to that depicted
in Fig.~\ref{aoki_crude}. Early numerical evidence supporting this
quenched phase structure may be found in Ref.~\onlinecite{aokiq}.
The weak-coupling region may also be studied via
an effective lagrangian \cite{GSS}.

As far as DWF and overlap fermions are concerned, the requirement of
a gap in the Wilson spectrum should be replaced by the requirement
that $\lambda_c>0$ \cite{lclz}.
In other words, one must work
outside of the (quenched) Aoki phase, in one of the C phases.
It is therefore important to map out the Aoki phase on any ensemble
used for DWF or overlap-fermion numerical simulations.
Furthermore, for practical reasons,
one should not be too close to the Aoki phase.
How close is ``too close'' depends on the underlying Boltzmann weight
and on the construction of the fermion operator.
We will discuss this very practical point at some length in our
conclusions.

In this paper we study the spectral properties of the Wilson operator $H_W$
via calculation of its resolvent $(H_W-z)^{-1}$ and correlation
functions derived from it.
The theoretical framework developed in Ref.~\onlinecite{lclz}
is directly applicable, and guides us in the numerical implementation.
We measure the spectral density
as well as properties that characterize the shape and size
of the localized eigenmodes.
The resolvent gives us these quantities much more economically than
would the direct study of the eigenvalues and eigenvectors of $H_W$.
The correlation functions address the Ward identities and the Banks--Casher
relation directly.

The resolvent allows simultaneous treatment of localized and extended
modes.
In any volume, the eigenvalues corresponding to localized modes are random.
When the resolvent is averaged over the gauge ensemble, the single-configuration
spectral density, $V^{-1}\sum_n\delta(\lambda-\lambda_n)$, is smeared.
Thus the ensemble-averaged spectral density is
a continuous function of the eigenvalue for localized as well as extended modes.
The essential physics of the localized modes lies not in their
discreteness but in the compactness of their wave functions.

Our measurements are carried out on
pure-gauge ensembles (which is the usual meaning of quenching).
We compare the spectral
properties for three different pure-gauge actions: the standard plaquette
action and
the Iwasaki \cite{iwa,iwaval} and DBW2 \cite{dbw2,dbw2val}
actions, two gauge actions motivated by renormalization-group considerations.
These gauge actions have been used in quenched \cite{iwCPPACS,dbw2RBC}
and dynamical \cite{ddwf} DWF simulations, and in quenched overlap simulations
\cite{1gev,gall}.

This paper is organized as follows. In Sec.~2 we give basic definitions
and derive the relevant Ward identities. In Sec.~3 we review
the Banks--Casher relation as well as the
localization alternative to Goldstone's theorem.
A twisted-mass term \cite{aoki,tm} provides the ``magnetic field''
that determines the direction of the pion condensate.
Careful study of the vanishing twisted-mass limit reveals that,
if the low-lying eigenmodes are localized, the two-point function of the
would-be Goldstone pions diverges linearly with the inverse twisted mass.
This enables the relevant Ward identity
to be saturated without a Goldstone pole.

We then turn to our numerical investigations, starting with the
standard plaquette action for the gauge field.
In Sec.~4 we present results for the simplest quantity, the spectral density.
In Sec.~5 we define the localization length and use it to
determine the mobility edge $\l_c$ for several points
$(g_0,m_0)$ in the phase diagram. Extrapolations of $\l_c$ to zero allow us
to map out a part of the boundary of the Aoki phase.
We then proceed to detailed study of the localized modes.
In Sec.~6 we extend the investigation to the Iwasaki and DBW2 gauge actions.
We conclude
in Sec.~7 with a discussion of the implications of our results for domain-wall
and overlap fermions.
The results of Sec.~5 yield several quantities that help locate regions of the phase diagram to be avoided in simulations.

A concise account of this work, not including the improved gauge actions, has already been given \cite{mob}.

\section{Definitions}
\label{Def}

\subsection{Fermion action}
\label{DefA}

The Wilson--Dirac operator is defined as
\begin{equation}
   D(m_0) =  {1\over a}\left(\begin{array}{cc}
      (W + a m_0)   & -C     \\
      C\dg  & (W + a m_0)
       \end{array}\right),
\label{DW}
\end{equation}
where
\begin{equation}
   C_{xy}  =  \half \sum_\m \left[\d_{x+\hat\m,y} U_{x\m}
    - \d_{x-\hat\m,y} U^\dagger_{y\m} \right] \s_\m
\end{equation}
comes from the naive Dirac operator, and
\begin{equation}
    W_{xy}  =  4\d_{xy} -\half \sum_\m \left[\d_{x+\hat\m,y} U_{x\m}
               + \d_{x-\hat\m,y} U^\dagger_{y\m} \right]
\end{equation}
is the Wilson operator that breaks chiral symmetry while
preventing species doubling. Here $\s_\m=(\bm\s,i)$, where $\s_k$
are the three Pauli matrices; we are using a chiral basis for the
Dirac matrices, where $\gamma_5$ is diagonal.  $U_{x\m}$ is the
SU($N$) matrix representing the gauge field. We study the spectrum
of the hermitian Wilson--Dirac operator,
\begin{equation}
  H_W(m_0)=D(m_0)\gamma_5.
\end{equation}
The corresponding eigenvalue equation (in a given gauge field) is
\begin{equation}
  H_W(m_0)\J_n=\l_n\J_n,
\label{Heigen}
\end{equation}
and we normalize the eigenvectors according to
$a^4\sum_x|\J_n(x)|^2=1$.

Previous studies of the spectrum of $H_W$ \cite{SCRI,JLSS,AT} were
based on the calculation of individual eigenfunctions and
eigenvalues. We find it more economical to calculate the Green
function
\begin{equation}
  G(z) = [H_W(m_0)-z]^{-1},
\label{G}
\end{equation}
where $z=\l+im_1$, in order to extract information about the spectrum.
$G(\l+im_1)$ is well defined in finite volume provided $m_1\ne 0$.
It has the spectral representation
\begin{equation}
  G(x,y;\l+im_1) = \sum_n
  {\J_n(x) \J_n^\dagger(y)\over \l_n - \l - im_1}.
\label{specrep}
\end{equation}

\subsection{Two flavors and twisted mass}

As mentioned in the Introduction, the spectral properties of $H_W$ have profound effects on the realization of continuous symmetries when there is more than one flavor.
Thus we will add an isospin index to the fermion field and
consider the two-flavor theory defined by
\begin{eqnarray}
  S_F &=&  \bj (H_W-\lambda)  \psi'
\nonumber\\[2pt]
  &=& \bj (D-\lambda\gamma_5) \psi,
\label{S0}
\end{eqnarray}
where $\psi'=\gamma_5\psi$.
Spontaneous breaking of the flavor symmetry (and of parity) will
be connected with the condensation of the ``pion'' field,
\begin{eqnarray}
\p_\pm(x)&=& i\bj(x)\g_5\t_\pm\j(x)\nonumber\\
\p_3(x)&=& i\bj(x)\g_5\t_3\j(x),
\label{pion}
\end{eqnarray}
where $\t_\pm=(\t_1\pm i\t_2)/2$.
The parameter $\lambda$ has been introduced into Eq.~(\ref{S0}) in order to shift the focus from zero to nonzero eigenvalues of $H_W$.
In order to control the isospin orientation of the condensate,
we add to the action a ``magnetic field'' in the guise of
a twisted-mass term, giving finally
\begin{eqnarray}
  S_F &=&  \bj [H_W-(\lambda+i\tau_3m_1)]  \psi'
\nonumber\\[2pt]
  &=& \bj [D-(\lambda+i\tau_3m_1)\gamma_5] \psi.
\label{S}
\end{eqnarray}
$m_1$ will be used as
a regulator to avoid the singularities of $G(z)$ along the real axis.

\subsection{SU(2) flavor symmetry and Ward identities}
\label{DefB}

For $m_1=0$, the fermion action~(\ref{S}) has a (vector) SU(2)
flavor symmetry.
For $m_1 \ne 0$, the Ward identity of the broken
symmetry is obtained by performing a local flavor transformation,
\begin{equation}
\psi(x)\to\psi(x)+\alpha(x)\delta_+\psi(x),
\end{equation}
and similarly for $\bar\psi(x)$, where
\begin{equation}
  \d_+\j(x)=i\t_+\j(x),\quad
  \d_+\bj(x)=-i\bj(x)\t_+ .
\label{ISO}
\end{equation}
We find for any operator $\co$ that
\begin{equation}
  \db_\m \svev{J^+_\m(x)\co(y)} + 2 m_1 \svev{\p_+(x)\co(y)}
    = {i\d_{xy}\over a^4} \svev{\d_+\co(y)}.
\label{gen}
\end{equation}
Here the backward lattice derivative is defined by $ \db_\m f(x) =
[ f(x) - f(x-\hat\m)]/a$, and the vector current corresponding to
Eq.~(\ref{ISO}) is
\begin{equation}
  J^+_\m(x) = \half \left[\bj(x) \t_+(\g_\m-1) U_\m(x)\j(x+\hat\m)
  + \bj(x+\hat\m) \t_+(\g_\m+1) U_\m^\dagger(x)\j(x) \right] .
\label{J}
\end{equation}
While the notation $\langle\dots\rangle$ indicates an integration
over both fermion and gauge fields, the Ward identity (\ref{gen})
in its various guises is in fact valid for each gauge
configuration separately.

We define the ``pion'' two-point function
\begin{equation}
  \G(x,y)=\svev{\p_+(x)\p_-(y)}
\label{Gammas}
\end{equation}
as well as
\begin{equation}
  \G_\m(x,y)=\svev{J_\m^+(x)\p_-(y)}.
\end{equation}
[Note that this is the usual pion only when $\lambda=m_1=0$; see
Eq.~(\ref{S}).]
If we take $\co(y)=\p_-(y)$ in Eq.~(\ref{gen}), we find the Ward identity
\begin{equation}
  \db_\m\G_\m(x,y) + 2 m_1\G(x,y)={\d_{xy}\over a^4}
  \svev{\p_3(y)}.
\label{pp}
\end{equation}
Applying a Fourier transform, {\em viz.}
\begin{eqnarray}
  \tG(p) &=& {a^8\over V} \sum_{xy} e^{ip(y-x)}\G(x,y)
\label{Gammap}\\
  \tG_\m(p) &=& {a^8\over V} \sum_{xy} e^{ip(y-x)}\G_\m(x,y),
\label{Gammamup}
\end{eqnarray}
we derive the momentum-space Ward identity,
\begin{equation}
  {1\over a} \sum_\m (1-e^{-iap_\m}) \tG_\m(p) + 2m_1 \tG(p)
  = \svev{\p_3}.
\label{ppft}
\end{equation}

\section{Goldstone's Theorem and localization}
\label{Gold}

The Ward identity~(\ref{ppft}) is valid for arbitrary $\l$ and $m_1$.
In a quenched theory, however, despite the
Goldstone Theorem, $\svev{\p_3} \ne 0$
does {\it not} necessarily imply the existence of a massless pole
in $\tG_\m(p)$ in the limit $m_1\to 0$.
Let us recall \cite{lclz,ms} how this comes about.

\subsection{Condensate and Banks--Casher relation}
\label{GoldA}

The volume-averaged pion condensate
\begin{equation}
\label{V}
  \svev{\p_3} = (a^4/V) \sum_x \svev{\p_3(x)},
\end{equation}
in the two-flavor
theory~(\ref{S}) can be expressed in terms of the Green function
$G(\lambda\pm im_1)$. We will denote the expectation value in a given gauge
field by $\langle\dots\rangle_\cu$. Then \begin{eqnarray} a^4 \sum_x
\svev{\p_3(x)}_\cu
    &=&-i a^4\,\Tr [G(\lambda+im_1)-G(\lambda-im_1)]\nonumber\\
  &=&  2\, \sum_n {m_1\over (\l_n-\l)^2  + m_1^2}\;,
\label{aoki}
\end{eqnarray}
where we have used the spectral representation (\ref{specrep}).
Averaging this over the gauge field gives the
translation-invariant result
\begin{equation}
   \svev{\p_3} = 2 \int d\l'\, \rho(\l')  {m_1\over (\l'-\l)^2+m_1^2} \,,
\label{pBC}
\end{equation}
where $\rho(\l)$ is the eigenvalue density defined by
\begin{equation}
  \r(\l) = {1\over V} \vev{\sum_n \d(\l-\l_n)} .
\label{rho}
\end{equation}
In the limit $m_1\to 0$, we obtain
\begin{equation}
  \svev{\p_3} = 2\p \r(\l).
\label{bc}
\end{equation}
This is a generalized Banks--Casher relation; the original relation
\cite{BC} is Eq.~(\ref{bc}) at $\lambda=0$.

\subsection{Localization as an alternative to Goldstone's theorem}
\label{GoldB}

Naively taking the limit $m_1\to 0$ in Eq.~(\ref{ppft}) gives
\begin{eqnarray}
  \svev{\p_3} &\stackrel{?}{=}& {1\over a} \sum_\m (1-e^{-iap_\m}) \tG_\m(p)\nonumber\\
  &\approx& ip_\m \tG_\m(p),
\label{false}
\end{eqnarray}
where the second line is the approximate form for $ap\ll 1$.
As we shall see shortly, Eq.~(\ref{false}) is sometimes false, but
it contains the Goldstone Theorem:
$\tG_\m(p)$ must have a massless pole
for any $\l$ such that $\svev{\p_3}\ne0$.
By the generalized Banks--Casher relation~(\ref{bc}),
this happens whenever $\r(\l) \ne 0$. Apart from  $\G_\m(x,y)$,
one expects long-range power-law decay also for other
correlation functions, including in particular $\G(x,y)$.

In the physics of disordered systems \cite{DJT}
it is well known that the eigenmodes of a hamiltonian in a random background
divide into two classes: {\it extended} and {\it localized}.
In fact, the spectrum splits into bands,
each band containing only eigenmodes of one type.%
\footnote{There seems to be no rigorous proof of this
fact except in one dimension \cite{DJT}.}
  A point in the spectrum separating an extended band from
a localized band is  a {\em mobility edge}.
$\G_\m(x,y)$ will exhibit a power-law decay when the $\l$ lies in
an extended band,
while for a localized band it will decay exponentially.
We expect that the same basic separation applies to $H_W$ as well.

  If $\r(\l)$ comes from localized eigenmodes and $\tG_\m(p)$
has no pole at zero momentum, what has become of the Goldstone Theorem?  In other words, how is the Ward identity~(\ref{ppft})
satisfied? The way out of this conundrum is the following.
In the limit $m_1\to 0$, $\tG(p)$ {\it diverges} as $1/m_1$
for a range of values of $p$ that includes the point $p=0$.
The limiting value of  $m_1\tG(p)$ is finite.
For $p\to0$, we arrive at an alternative
to Eq.~(\ref{false}),
\begin{equation}
\lim_{m_1\to0}2m_1\tG(0)=\lim_{m_1\to0}\svev{\p_3}.
\label{altr}
\end{equation}

\subsection{Divergence of the pion two-point function}
\label{GoldC}

Let us consider further the $1/m_1$ divergence in $\tG(p)$.
The (finite-volume) spectral representation of the charged-pion
two-point function is
\begin{widetext}
\begin{equation}
  \G(x,y) =
  \Bigl\langle
    \sum_{n^{\pm}}
    \J_{n^+}^\dagger(x) \J_{n^-}(x) {1\over \l_{n^-} - \l + im_1}
    \J_{n^-}^\dagger(y) \J_{n^+}(y) {1\over \l_{n^+} - \l - im_1}
  \Bigr\rangle,
\label{dblsum}
\end{equation}
\end{widetext}
where terms with the subscript
$n^{\pm}$ are associated with the propagator for the
corresponding quark flavor. As explained in Sec.~3 of Ref.~\onlinecite{lclz},
a $1/m_1$ divergence may arise only from the terms with $n^+=n^-$,
so that
\begin{equation}
  \G(x,y) = {1\over m_1}
  \biggl\langle
    \sum_n |\J_n(x)|^2 |\J_n(y)|^2 {m_1 \over (\l_n - \l)^2 + m_1^2}
  \biggr\rangle + O(1).
\label{limm1}
\end{equation}
In analogy with Eq.~(\ref{rho}), we define the
eigenmode-density correlation function,
\begin{equation}
  \car_\l(x,y) =
  \Bigl\langle
    \sum_n  |\J_n(x)|^2 |\J_n(y)|^2  \d(\l-\l_n)
  \Bigr\rangle,
\label{Rdef}
\end{equation}
and its Fourier transform,
\begin{equation}
  \tR_\l(p) = {1\over V}\Bigl\langle
    \sum_n  |H_n(p)|^2 \, \d(\l-\l_n)
  \Bigr\rangle,
\label{tRdef}
\end{equation}
where
\begin{equation}
  H_n(p) = a^4\sum_x|\J_n(x)|^2 e^{ipx} .
\label{Hp}
\end{equation}
As $m_1\to0$, these may be interpreted as the contribution to $\G(x,y)$,
or to its Fourier transform, of eigenmodes with eigenvalue $\lambda$. Repeating
the analysis leading from Eq.~(\ref{aoki}) to Eq.~(\ref{bc}) we find, for
$m_1 \ll 1$,
\begin{equation}
  \G(x,y) = {\p \car_\l(x,y)\over m_1}  +  O(1) .
\label{Rm1}
\end{equation}
Observe that $\car_\l(x,y)$ is the ensemble average~(\ref{Rdef})
of a quantity that is strictly positive. Also, since
the eigenmodes are normalized, we have that
\begin{equation}
  \tR_\l(0) = {a^8\over V}\, \sum_{xy} \car_\l(x,y) = \r(\l) ,
\label{Rrho}
\end{equation}
and so $\r(\l) \ne 0$ means that $\car_\l(x,y)$
must be nonzero for at least one pair of values $(x,y)$.
Consequently, in any \textit{finite} volume, $\r(\l) \ne 0$ implies the
existence of a $1/m_1$ divergence in the coordinate-space two-point function.
[This must be true for at least one value of $(x,y)$,
but is expected to hold for practically every $(x,y)$.] This
result is valid for the generalized quenched theory defined by
Eq.~(\ref{S}) for any Boltzmann weight; the only assumption we used is
that the Boltzmann weight does not depend on $\l$ and $m_1$.

  [An {\it unquenched} theory with fermion action~(\ref{S})
would include ${\rm det}(H_W - \lambda -i\tau_3 m_1)$
in the Boltzmann weight.
This determinant will suppress
eigenvalues $\l_n \approx \l$, and the spectral density
measured by Eq.~(\ref{bc}) will be zero in any finite volume in the limit $m_1\to0$.]

We next consider the infinite-volume limit.
The asymptotic behavior of an exponentially localized eigenmode is
\begin{equation}
  | \J_n(x) |^2 \sim
  \exp\left(- {|x-x_n^0| \over l_n} \right) ,
  \qquad |x-x_n^0| \gg l_n ,
\label{lclx}
\end{equation}
where $l_n$ is the {\em localization length\/}
and $x_n^0$ is the center of the localized eigenmode.
Eq.~(\ref{lclx}) is valid only at distances that are large compared to
the size of the region containing most of the eigenmode's density.
In principle, nothing forbids the occurrence of
eigenmodes with a very short localization length $l_n \le a$.

An extended eigenmode is one that does not satisfy Eq.~(\ref{lclx})
for any finite $l_n$. Evidently, truly extended eigenmodes
exist only in infinite volume. In finite
volume, the clear-cut identification of an eigenmode as localized
demands that
$|\J_n(x)|^2$ is exponentially small on most of the lattice.
In later sections we will give a more quantitative criterion.

In infinite volume, we expect the Fourier transform
of the eigenmode density (\ref{lclx}) to have
the following small-$p$ behavior:
\begin{equation}
  H_n(p) \approx {e^{ipx_n^0}\over 1 + p^2 l_n^2} .
\label{Hn}
\end{equation}
The region $p^2 l_n^2 \ll 1$ will reflect
only the exponentially decaying envelope~(\ref{lclx}) of the eigenmode density
but not the short-distance fluctuations.
The overall normalization is set by $H_n(0)=1$ for a normalized eigenmode.
Substituting the {\it ansatz}~(\ref{Hn}) into Eq.~(\ref{tRdef})
and going to small $p$ we find, in analogy with Eq.~(\ref{Rm1}),
\begin{equation}
  \tG(p) = {\p\r(0)\over m_1}
  \left[1 + O\! \left( p^2\, \overline{l}^2 \right)\right] +O(1),
\label{noGB}
\end{equation}
where $\overline{l}$ is some average localization length
for the eigenmodes with eigenvalue $\l$.
[This extends Eq.~(\ref{altr}) to nonzero $p$.]
{}From Eq.~(\ref{ppft}) we conclude that
\begin{equation}
\tG_\m(p) \sim ip_\m \r(0)\overline{l}^2.
\end{equation}
Thus there is no Goldstone pole when the eigenmodes with eigenvalue $\l$
are exponentially localized.\footnote{See also
Sec.~4 of Ref.~\onlinecite{lclz}.}

When the eigenmodes at the given $\l$ are extended,
the transition from Eq.~(\ref{dblsum}) to Eq.~(\ref{limm1}) is not justified
if the limit $m_1\to 0$ is taken after the infinite-volume limit,
because of interference effects between eigenmodes with
infinitesimally close eigenvalues.
In this case we expect Eq.~(\ref{false}) to be valid,
indicating a Goldstone pole.

\section{Wilson plaquette action: spectral density}
\label{Rho}

We now turn to our numerical investigation.
We begin with quenched ensembles generated with the Wilson
plaquette action for the SU(3) gauge theory,
\begin{equation}
S=\frac\beta3\sum_{x\atop\mu<\nu}{\rm Re}\,\Tr (1-U_{x\mu\nu}).
\label{plaq}
\end{equation}
We began calculating at $\beta=6.0$, which is usually taken to correspond to
the lattice scale $a^{-1}\simeq2$~GeV, and at%
\footnote{For the remainder of the paper we rescale $am_0\to m_0$,
giving the bare Wilson mass in lattice units.}
 $m_0=-1.5$,
between the Aoki ``fingers'' that point to the $\beta=\infty$
($g_0=0$) axis at $m_0=0$ and~$-2$.
Then we moved downward in $\beta$ towards the Aoki phase, calculating
at $\beta=5.85$, 5.7~(where $a^{-1}\simeq1$~GeV), 5.6, 5.5, and~5.4.
We will show that the Aoki phase is entered just below $\beta=5.6$
(see Fig.~\ref{aoki_crude}).

Returning to $\beta=5.7$, we moved ``sideways'' by changing $m_0$ to~$-2.0$
and $-2.4$. The latter turns out to be very close to or in the second Aoki finger.
We put these choices of $(\b,m_0)$ into the context of other work
in Sec.~\ref{DWF}.

For each value of $\beta$ we generated 120 uncorrelated gauge
configurations (except where otherwise noted) on a lattice of $16^4$ sites, using the MILC pure-gauge
overrelaxation code.
For a given $\beta$,
results for all values of $m_0$, $\lambda$, and~$m_1$ were calculated on
the same ensemble; thus correlations had to be taken into account in
all fits and statistical analysis.

We shall illustrate our methods by discussing in detail
the analysis for $\beta=5.85$ and $m_0=-1.5$.
Results for other
values of $(\beta,m_0)$ will be summarized in the tables.

\subsection{Spectral density from Banks--Casher relation}
\label{RhoA}

{}From the Banks--Casher relation (\ref{bc}) and Eq.~(\ref{aoki}) we have
\begin{equation}
  2\pi\rho(\lambda) = \lim_{m_1\to0}\svev{\pi_3}
  = {2\over V}\lim_{m_1\to 0} \svev{{\rm Im}\, \Tr G(\lambda+im_1)}.
\label{unrho}
\end{equation}
Thus the volume- and ensemble-averaged Green function,
extrapolated to $m_1=0$, gives $\rho(\lambda)$ directly. Of course
$m_1$ must be kept nonzero for actual calculation in order for $G$
to be bounded.

The spectral sum~(\ref{aoki}) shows how to do the extrapolation.
For any gauge configuration $\cal U$ we consider the sum
\begin{equation}
  \sum_n {m_1\over (\l_n-\l)^2  + m_1^2} .
\label{specsum}
\end{equation}
The summand tends to a $\delta$-function as $m_1\to 0$, but before
the limit is taken it has a finite width equal to $m_1$. The given
configuration will make a contribution of $O(1/m_1)$ if $H_W$ has
an eigenmode whose eigenvalue $\lambda_n$ satisfies
$|\lambda_n-\lambda|\alt m_1$; these contributions, summed over
configurations $\cal U$, will add up to a finite limit as
$m_1\to0$. On the other hand, all the eigenmodes that are far from
$\lambda$, with $a|\lambda_n-\lambda|=O(1)$, will make a
contribution of $O(m_1)$. This indicates a linear extrapolation,
\begin{equation}
  {\langle\pi_3\rangle\over 2\pi} = c_0 + c_1m_1  ,
\label{rfit}
\end{equation}
where $c_{0,1}$ will depend on $\l$.
Then $c_0$ is an estimate for $\r(\l)$.

\begin{figure}[htb]
\begin{center}
\includegraphics*[width=.7\columnwidth]{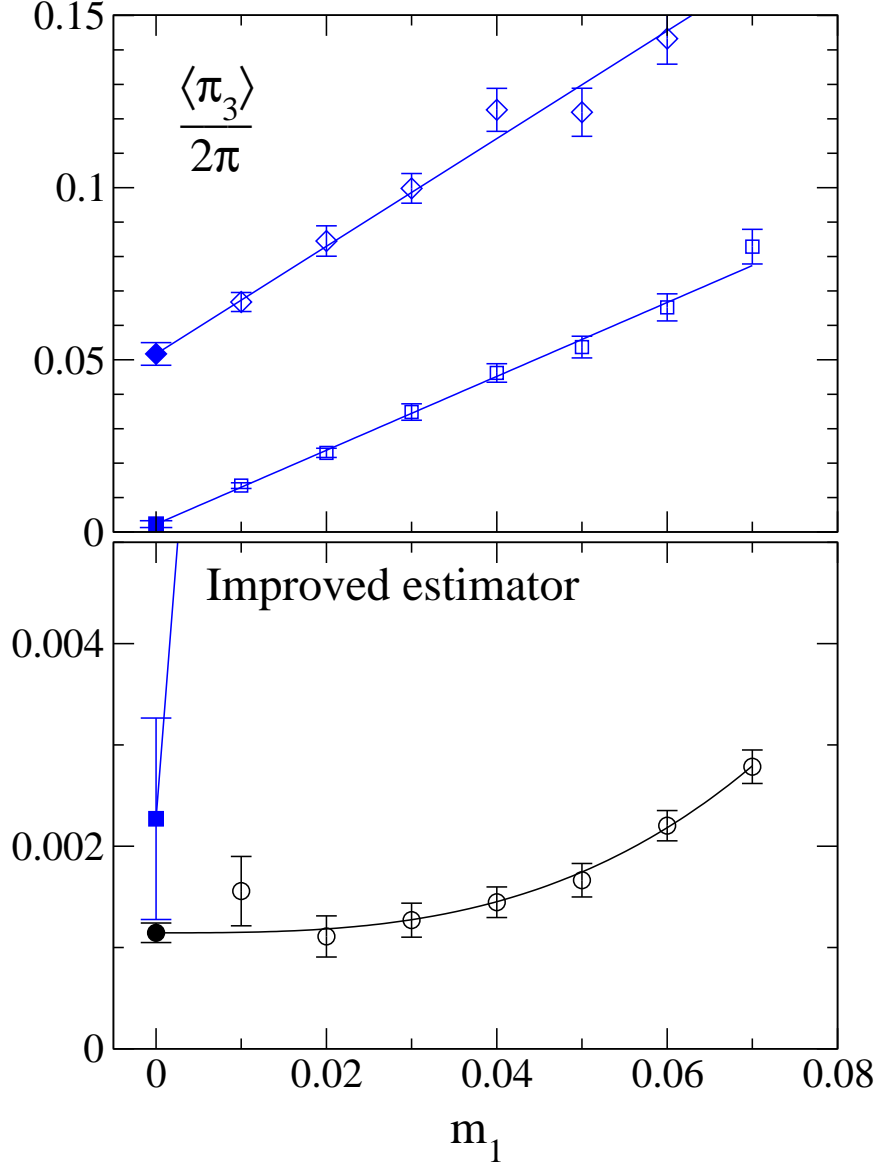}
\caption{Top graph: the pion condensate
$(2\pi)^{-1}\langle\pi_3\rangle$ as a function of $m_1$ (open
symbols), and its linear extrapolation to yield the spectral
density $\rho(\lambda)$ at $m_1=0$, for $\lambda=0.3$ (upper
curve) and $\lambda=0$ (lower curve). Bottom graph: cubic
extrapolation of improved estimator for $\lambda=0$ (circles),
compared to the result of the linear fit above (square). All data
are for Wilson action at $\beta=5.85$; $m_0=-1.5$.
\label{b5.85_l0.0_rho_vs_m1}}
\end{center}
\end{figure}

We calculated $\Tr G$ using a single random source per gauge configuration.
Averages were obtained for up to seven values of
$m_1$: 0.01, 0.02, $\ldots$, 0.07.
The upper graph in Fig.~\ref{b5.85_l0.0_rho_vs_m1} shows
the linear extrapolation for two values of $\lambda$.
For $\lambda=0.3$ the fit works well.
For $\l=0.$, on the other hand, the extrapolated intercept
is very small and the precision attained is inadequate.

The problem is that, when $\r(\l)$ is too small,
there are few eigenmodes (for the volume we use) within the broadened $\delta$-function in
Eq.~(\ref{specsum}).
Then the sum is dominated by the $O(m_1)$ contributions of the more
distant eigenmodes, with large fluctuations.
Making $m_1$ even smaller will suppress these contributions, but
it will be self-defeating because even fewer modes will lie within
the broadened $\delta$-function, increasing the fluctuations in their
contribution as well.

\subsection{Improved estimator}
\label{RhoB}

A solution lies in using an improved estimator for the spectral
density, one that suppresses the contribution of distant eigenmodes
and thus approaches the $m_1\to 0$ limit faster than linearly.
We introduce the (dimensionless) differential operator $\cj$ defined by
\begin{equation}
   \cj f = -m_1^2 \; {\partial \over \partial m_1}
   \left( {1\over m_1} f \right)
\label{diffm}
\end{equation}
for any $f(m_1)$.
It is designed to remove the linear term in $f(m_1)$.
Applying it to the pion condensate, we have
\begin{equation}
  \cj \svev{\p_3}
  = {2\over V}\svev{{\rm Im}\, \Tr G - m_1 \,{\rm Re}\, \Tr G^2},
\label{imppi}
\end{equation}
with the spectral representation
\begin{equation}
  \cj \svev{\p_3}= {4\over V}\biggl\langle{\sum_n {m_1^3 \over [(\l_n-\l)^2 + m_1^2]^2}}\biggr\rangle.
\label{imprho}
\end{equation}
The $m_1\to 0$ limit again yields $2\pi\r(\l)$ since
$\int dx\, m_1^3/(x^2 + m_1^2)^2=\p/2$.

The operator $\cj$ indeed removes
the leading, linear contribution of the eigenmodes with $a|\l_n-\l|=O(1)$.
We can see from Eq.~(\ref{imprho}) that
the contribution of these eigenmodes is now proportional to $m_1^3$.
In view of this, we attempt a {\em cubic\/} extrapolation,
\begin{equation}
  {\cj\langle\pi_3\rangle\over 2\pi} = c_0 + c_3 m_1^3 .
\label{cubfit}
\end{equation}
The lower graph in Fig.~\ref{b5.85_l0.0_rho_vs_m1} shows the new extrapolation,
again at $\l=0$. Linear and quadratic terms are in fact unnecessary to attain an excellent fit to the data points; this is the final justification of the model (\ref{cubfit}).
Most important, the precision in the extrapolation is improved by a factor of ten.

Using the improved estimator~(\ref{imppi}) requires a second inversion
for each random source. The number of CG iterations needed for the second
inversion can reach twice the number for the first inversion.
The cost of the improved estimator is thus up to (roughly) three times
the original cost. Were we to invest the additional computer time
instead in increased statistics
using the simple estimator~(\ref{unrho}), the anticipated reduction
in the statistical error would not come anywhere close to the factor
of ten achieved with the improved estimator.

\subsection{Spectral density from the two-point function}
\label{RhoC}
Since the two-point function $\G(x,y)$ is the eigenmode-density correlator
[see Eq.~(\ref{limm1})], it
is a rich source of information about spectral properties.
The first application we discuss is an alternative method to calculate
the spectral density $\rho(\lambda)$.

We project $\G(x,y)$ to zero spatial momentum and calculate the
time-correlation function,
\begin{equation}
\G(t)={a^6\over V_3}\sum_{\mathbf{x},\mathbf{y}}\G(0,\mathbf{x};t,\mathbf{y}),
\label{Gt}
\end{equation}
where $V_3$ is the three-volume.
This calculation requires a random source on time slice 0,
as well as a random sink for each time slice $t$;
again we use one set of random sources per gauge configuration.
If we sum $\G(t)$ over all $t$, we can use translation invariance
and Eq.~(\ref{limm1}) to show that
\begin{equation}
  \lim_{m_1\to 0} am_1 \sum_t \G(t) = \p\r(\l) .
\label{sumGt}
\end{equation}
This corresponds to setting $p=0$ in the
finite-volume Ward identity~(\ref{ppft}).
[Note the similarity to Eq.~(\ref{Rrho}).]
We verified Eq.~(\ref{sumGt}) for two values of the bare
coupling ($\b=6.0$ and~5.7) by calculating
both the condensate [Eq.~(\ref{unrho})] and the two-point function.
In all subsequent measurements we did not calculate the condensate
separately since it gives only the spectral density, while $\Gamma(t)$ yields
this and much more.

The extrapolation of $m_1\G(t)$ to $m_1=0$ suffers from the same fluctuation
problems as that of $\langle\pi_3\rangle$ when the spectral density is
small.
The solution again lies in an improved estimator,
but now for the eigenmode-density
correlation function $\car(x,y;\l)$ [see Eq.~(\ref{Rdef})].
We define the ``improved'' two-point function,
\begin{equation}
  \Gimp(x,y) = \Tr \svev{\cg(x,y)\cg(y,x)},
\label{impGx}
\end{equation}
where
\begin{eqnarray}
  \cg(x,y) &=&  a^4 \sum_z G(x,z;\l+im_1) G(z,y;\l-im_1)\nonumber\\
  &=& \sum_n {\J_n(x) \J_n^\dagger(y)\over (\l_n - \l)^2 + m_1^2},
\label{GG}
\end{eqnarray}
and project it to zero spatial momentum,
\begin{equation}
  \Gimp(t)
  = {a^6\over V_3} \sum_{\mathbf{x},\mathbf{y}} \Gimp(0,\mathbf{x};t,\mathbf{y}).
\label{impGt}
\end{equation}
Instead of Eq.~(\ref{limm1}) one has
\begin{widetext}
\begin{equation}
  m_1^3\,\Gimp(x,y) = \biggl\langle
    \sum_n |\J_n(x)|^2 |\J_n(y)|^2 {m_1^3 \over [(\l_n - \l)^2 + m_1^2]^2}
  \biggr\rangle + O(m_1^2),
\label{impGm1}
\end{equation}
\end{widetext}
whence
\begin{equation}
  \lim_{m_1\to 0} m_1^3\,\Gimp(x,y) = {\p\over 2} \car_\l(x,y).
\label{impGlim}
\end{equation}
Equation~(\ref{Rrho}) then yields the spectral density [cf.~Eq.~(\ref{sumGt})],
\begin{equation}
  \lim_{m_1\to 0} a m_1^3 \sum_t \Gimp(t) = {\p \over 2}\r(\l).
\label{sumGimp}
\end{equation}
Results obtained using $\G(t)$ were extrapolated to $m_1=0$
linearly in $m_1$,
while those
obtained using $\Gimp(t)$ were extrapolated%
\footnote{The form of Eq.~(\ref{impGm1}) suggests an $m_1^2$ term as well,
but we found it to be unnecessary for a good fit.}
linearly in $m_1^3$.

We also tried to improve the estimator for the eigenmode density
by calculating $\cj\G(x,y)$ [cf.~Eq.~(\ref{diffm})].
This has the advantage that the application of $\cj$ to all terms in the
Ward identity~(\ref{ppft}) generates a new identity.
As it turns out, while some reduction in the statistical error
was achieved this way, the improvement was not significant.
The reason is that the application of $\cj$ to $\G(x,y)$
would be effective in suppressing the contribution of eigenmodes
with $a|\l_{n^\pm}-\l|\approx 1$
only if the approximation Eq.~(\ref{limm1}) is valid.
As it turns out, for the $m_1$ values we used,
Eq.~(\ref{limm1}) is not a good approximation to Eq.~(\ref{dblsum})
when the spectral density is small.
The correlator $\Gimp(x,y)$ does a better job in suppressing
the contribution of eigenmodes with $a|\l_{n^\pm}-\l|\approx 1$.

\subsection{Numerical results}
\label{RhoD}
\begin{figure}[htb]
\begin{center}
\includegraphics*[width=.7\columnwidth]{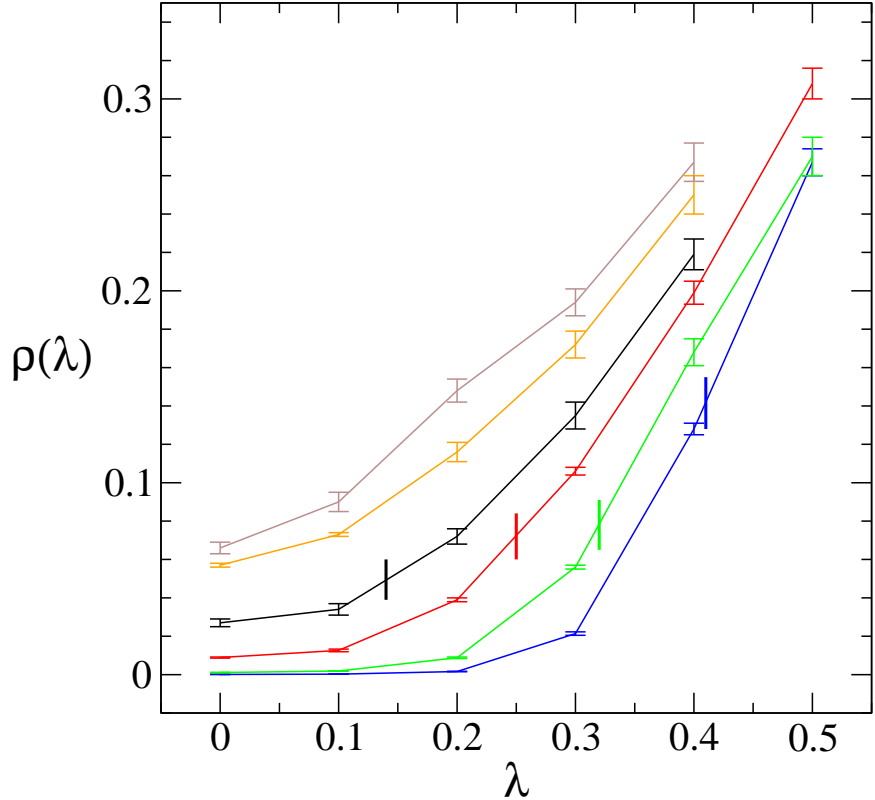}
\caption{Eigenvalue density $\rho(\lambda)$ for (bottom to top)
$\beta=6.0$, 5.85, 5.7, 5.6, 5.5, 5.4.
Mobility edges $\lambda_c(\beta)$ (when $\lambda_c>0$) are indicated by vertical bars.
Wilson action; $m_0=-1.5$.
\label{rho_vs_l_all_beta_linear}}
\end{center}
\end{figure}

\begin{figure}[htb]
\begin{center}
\includegraphics*[width=.7\columnwidth]{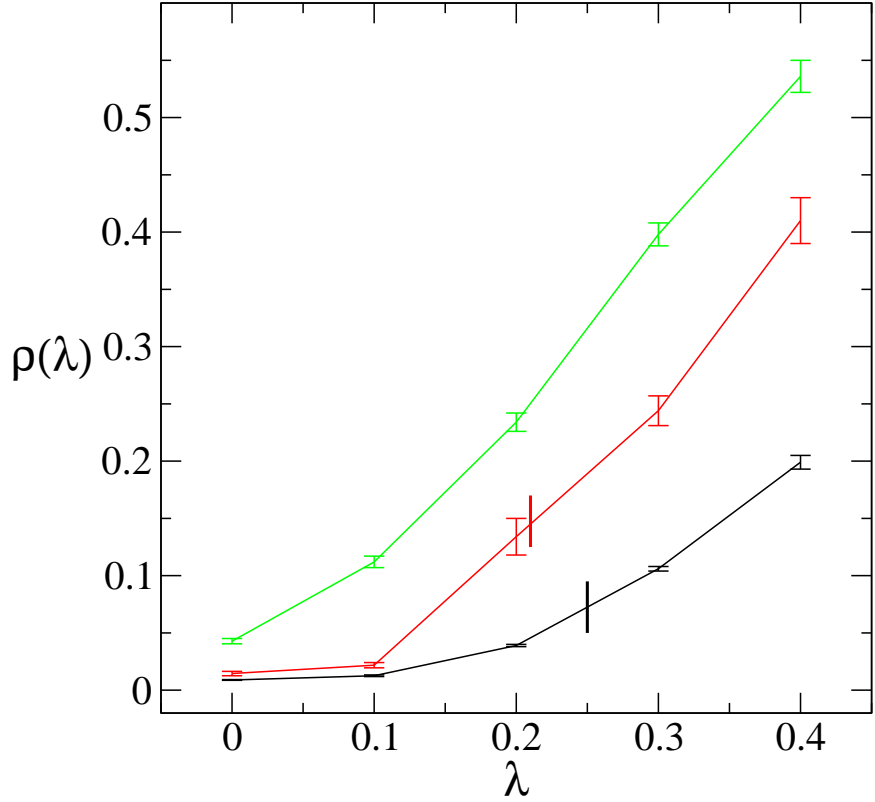}
\caption{Eigenvalue density $\rho(\lambda)$ for $\beta=5.7$ and (bottom to top)
$m_0=-1.5$, -2.0, -2.4.
Mobility edges are indicated by vertical bars.
For $m_0=-2.4$ the mobility
edge is probably very close to $\lambda=0$.  Wilson action.
\label{rho_vs_l_all_kappa_linear}}
\end{center}
\end{figure}

We present our results for the spectral density
for the Wilson plaquette action in Tables 1--8 and in Figs.~\ref{rho_vs_l_all_beta_linear} and~\ref{rho_vs_l_all_kappa_linear}.

At each value of $\beta$ and $m_0$, we performed our calculations
for a range of values of $\lambda$ starting from zero, increasing
in steps of~0.1. In the ranges studied (the maximal $\lambda$ was
0.4~to~0.6) the spectral density typically increases by one to three
orders of magnitude.
For the ensemble generated by the plaquette action, $\rho(\lambda)$ shows no
remarkable behavior as $\lambda$ passes $\lambda_c$, the mobility
edge (to be determined below).
In all cases, however, the mobility edge is encountered
for $\rho(\lambda_c)\approx 0.1$.
There is a steady rise in the spectral density as we decrease $\beta$
(Fig.~\ref{rho_vs_l_all_beta_linear}), which bespeaks an increased
disorder in the gauge field.

A final note on improvement:
For the larger values of $\l$ it is possible to achieve acceptable precision
without using the improved estimator. Since the number of
CG iterations required grows rapidly with the spectral density,
we limited the use of improved estimators to those cases where
results obtained without improvement were poor.

\begin{table*}[hb]
\caption{%
Spectral properties for $\beta=6.0$, $m_0=-1.5$. The mobility edge
is at $\lambda_c\simeq0.41$, marked by the horizontal line in the table.
The quantities $l_\ell$, $\cd$, $l_s$, and~$R$, which characterize the
localized modes, are defined in Sec.~\ref{Wil}.
\label{tab6.0}}
\begin{ruledtabular}
\begin{tabular}{dddddd}
\lambda & \rho & l_\ell & \cd & l_s & R  \\
\hline\hline
0.0\footnotemark[1]\footnotemark[2] &  0.00011(2)  &  0.61(1)  &  0.000038(8)  &  2.9(8)  &  31.   \\
0.1\footnotemark[1]\footnotemark[2] &  0.00024(2)  &  0.61(1)  &  0.00010(1)   &  2.4(3) &  25.   \\
0.2\footnotemark[1] &  0.0016(1)   &  0.70(1)  &  0.00053(6)   &  3.0(4) &  16.   \\
0.3\footnotemark[1] &  0.0214(9)   &  1.16(4)  &  0.0058(4)    &  3.7(3) &   8.   \\
0.4\footnotemark[1] &  0.128(3)    &  3.8(6)   &  0.025(2)     &  5.1(4) &   5.3  \\
\hline
0.5 &  0.267(7)    &  \infty   &  0.033(5)     &  8.(1)  &   4.4  \\
0.6 &  0.390(9)    &  \infty   &  0.042(9)     &  9.(2)  &   4.0  \\
\end{tabular}
\end{ruledtabular}
\footnotetext[1]{Results from improved estimator.}
\footnotetext[2]{Used 1200 gauge configurations for measurements with
$m_1=0.01$, 0.02, 0.03.}
\end{table*}

\begin{table*}[ht]
\caption{%
Spectral properties for $\beta=5.85$, $m_0=-1.5$. The mobility edge
is at $\lambda_c\simeq0.32$.
\label{tab5.85}}
\begin{ruledtabular}
\begin{tabular}{dddddd}
\lambda & \rho & l_\ell & \cd & l_s & R  \\
\hline\hline
0.0\footnotemark[1] &  0.0011(1)  &  0.64(1)  &  0.00036(5)  &  3.1(5)  &  17.    \\
0.1\footnotemark[1] &  0.0019(1)  &  0.71(1)  &  0.00054(7)  &  3.5(5)  &  15.    \\
0.2\footnotemark[1] &  0.0088(4)  &  0.92(2)  &  0.0024(2)   &  3.7(3)  &  10.    \\
0.3\footnotemark[1] &  0.056(1)   &  2.2(2)   &  0.0132(8)   &  4.2(3)  &  6.5    \\
\hline
0.4 &  0.168(7)   &  \infty   &  0.036(6)    &   4.7(8) &  4.9     \\
0.5 &  0.27(1)    &  \infty   &  0.023(7)    &  12.(4)  &  4.4     \\
0.6 &  0.39(2)    &  \infty   &  0.055(9)    &   7.(1)  &  4.0     \\
\end{tabular}
\end{ruledtabular}
\footnotetext[1]{Results from improved estimator.}
\end{table*}

\begin{table*}[ht]
\caption{%
Spectral properties for $\beta=5.7$, $m_0=-1.5$. The mobility edge
is at $\lambda_c\simeq0.25$.
\label{tab5.7}}
\begin{ruledtabular}
\begin{tabular}{dddddd}
\lambda & \rho & l_\ell & \cd & l_s & R  \\
\hline\hline
0.0\footnotemark[1] &  0.0089(3) &  1.04(3)  &  0.0023(2)  &  3.9(4)  &  10.   \\
0.1\footnotemark[1] &  0.0126(7) &  1.29(5)  &  0.0029(3)  &  4.3(5)  &   9.4  \\
0.2\footnotemark[1] &  0.039(1)  &  2.3(1)   &  0.0097(5)  &  4.0(2)  &   7.1  \\
\hline
0.3\footnotemark[1] &  0.106(2)  &  \infty   &  0.026(1)   &  4.1(2)  &   5.5  \\
0.4\footnotemark[1] &  0.199(6)  &  \infty   &  0.036(4)   &  5.5(6)  &   4.7  \\
0.5\footnotemark[1] &  0.308(8)  &  \infty   &  0.052(4)   &  5.9(5)  &   4.2  \\
\end{tabular}
\end{ruledtabular}
\footnotetext[1]{Results from improved estimator.}
\end{table*}

\begin{table}[ht]
\caption{%
Spectral properties for $\beta=5.6$, $m_0=-1.5$. The mobility edge
is at $\lambda_c=0.14(2)$. (We give the statistical error in $\lambda_c$ here
because it is larger than in other cases.  All suffer from
systematic error in the extrapolation.)
\label{tab5.6}}
\begin{ruledtabular}
\begin{tabular}{dddddd}
\lambda & \rho & l_\ell & \cd & l_s & R  \\
\hline\hline
0.0 &  0.027(2)  &  2.1(1)  &   -         &   -     &  7.8   \\
0.1 &  0.034(3)  &  4.(1)   &   -         &   -     &  7.4   \\
\hline
0.2 &  0.072(4)  &  \infty  &   0.012(3)  &   6.(2) &  6.1   \\
0.3 &  0.135(7)  &  \infty  &   0.016(5)  &   8.(3) &  5.2   \\
0.4 &  0.219(8)  &  \infty  &   0.017(6)  &  13.(5) &  4.6   \\
\end{tabular}
\end{ruledtabular}
\end{table}

\begin{table}[ht]
\caption{%
Spectral properties for $\beta=5.5$, $m_0=-1.5$. The mobility edge
is at zero.
\label{tab5.5}}
\begin{ruledtabular}
\begin{tabular}{dddddd}
\lambda & \rho & l_\ell & \cd & l_s & R  \\
\hline\hline
0.0\footnotemark[1] &  0.057(1)  &  \infty  &   0.0112(6) &   5.1(3) &   6.5   \\
0.1\footnotemark[1] &  0.073(1)  &  \infty  &   0.013(1)  &   5.6(4) &   6.1   \\
0.2 &  0.116(5)  &  \infty  &   0.015(4)  &   8.(2)  &   5.4   \\
0.3 &  0.172(7)  &  \infty  &   0.019(4)  &   9.(2)  &   4.9   \\
0.4 &  0.25(1)   &  \infty  &   0.029(6)  &   9.(2)  &   4.5   \\
\end{tabular}
\end{ruledtabular}
\footnotetext[1]{Results from improved estimator.}
\end{table}

\begin{table}[ht]
\caption{%
Spectral properties for $\beta=5.4$, $m_0=-1.5$. The mobility edge
is at zero.
\label{tab5.4}}
\begin{ruledtabular}
\begin{tabular}{dddddd}
\lambda & \rho & l_\ell & \cd & l_s & R  \\
\hline
0.0 &  0.066(3)  &  \infty  &  -         &  -      &  6.2   \\
0.1 &  0.089(5)  &  \infty  &  0.008(3)  &  11.(4) &  5.8   \\
0.2 &  0.148(6)  &  \infty  &  0.022(4)  &   7.(1) &  5.1   \\
0.3 &  0.194(7)  &  \infty  &  0.018(4)  &  11.(3) &  4.8   \\
0.4 &  0.27(1)   &  \infty  &  0.028(6)  &  10.(2) &  4.4   \\
\end{tabular}
\end{ruledtabular}
\end{table}

\clearpage
\pagebreak

\begin{table}[ht]
\caption{%
Spectral properties for $\beta=5.7$, $m_0=-2.0$. The mobility edge
is at $\lambda_c\simeq0.21$.
\label{tab5.7M2.0}}
\begin{ruledtabular}
\begin{tabular}{dddddd}
\lambda & \rho & l_\ell & \cd & l_s & R  \\
\hline\hline
0.0 &  0.014(2)  &  1.14(4)  &  0.005(2)  &  3.(1) &  9.1   \\
0.1 &  0.022(2)  &  1.37(5)  &  0.005(2)  &  4.(2) &  8.2   \\
0.2 &  0.13(2)   &  3.8(5)   &  0.03(1)   &  4.(2) &  5.3   \\
\hline
0.3 &  0.24(1)   &  \infty   &  0.028(4)  &  9.(1) &  4.5   \\
0.4 &  0.41(2)   &  \infty   &  0.05(1)   &  8.(2) &  4.0   \\
\end{tabular}
\end{ruledtabular}
\end{table}

\begin{table}[ht]
\caption{%
Spectral properties for $\beta=5.7$, $m_0=-2.4$. The mobility edge
is very close to zero.
\label{tab5.7M2.4}}
\begin{ruledtabular}
\begin{tabular}{dddddd}
\lambda & \rho & l_\ell & \cd & l_s & R  \\
\hline\hline
0.0 &  0.043(2)  &  4.(1)    &  0.006(2)  &   7.(2) &  6.9   \\
\hline
0.1 &  0.112(5)  &  \infty   &  0.014(3)  &   8.(2) &  5.5   \\
0.2 &  0.234(8)  &  \infty   &  0.026(6)  &   9.(2) &  4.5   \\
0.3 &  0.40(1)   &  \infty   &  0.052(6)  &   8.(1) &  4.0   \\
0.4 &  0.54(1)   &  \infty   &  0.05(1)   &  11.(2) &  3.7   \\
\end{tabular}
\end{ruledtabular}
\end{table}


\section{Wilson plaquette action: localization properties}
\label{Wil}

A central goal of the work presented here is the determination of the mobility edge $\lambda_c$ at various places in the $(\beta,m_0)$
phase diagram.
Equally interesting is the shape and density of the localized eigenmodes below $\lambda_c$ (where $\lambda_c>0$).

The two-point function $\G(x,y)$ contains complete information about the eigenmodes of $H_W$, and it is our primary tool both in determining $\lambda_c$ and in studying the localized modes.
When dealing with localized eigenmodes, we aim for parameter values
where the single spectral sum (\ref{limm1}) provides a good
approximation to the exact expression (\ref{dblsum}). When $m_1$ is small enough, $\G(x,y)$ then reduces to the
eigenmode-density correlator $\car_\l(x,y)$ [Eq.~(\ref{Rdef})].
Our analysis is based on this feature.
The approximation~(\ref{limm1}) ceases to hold when the eigenmodes become too dense or too extended, that is, when they interfere in the double sum.  This occurs when $\lambda$ is above or too close to the mobility edge.
We will develop a criterion to establish the consistency of our analysis of the localized eigenmodes.

We begin by giving a precise definition of the localization length $l_\ell(\lambda)$.  Its divergence marks the mobility edge $\lambda_c(\beta,m_0)$.
By definition, $\lambda_c=0$ marks the Aoki phase.
We then turn to study other properties of the localized eigenmodes
outside the Aoki phase, where $\lambda_c>0$.

\subsection{The localization length}
\label{WilA}
We have loosely defined the localization length $l_n$ of an individual eigenmode in Eq.~(\ref{lclx}).
We can use $l_n$, nebulous as it is, to motivate the definition of an {\em average\/} localization length that can be obtained from from the large-$t$ behavior of $\G(t)$.
Once we reach this definition, a precise definition of $l_n$ will be superfluous.

We begin by introducing the restricted spectral density $\r_l(\l)$ that
contains the contributions of eigenmodes with localization length
$l_n \le l$ only.  It is given by
\begin{equation}
  \r_l(\l) = {1\over V}\, \vev{\sum_n \d(\l_n-\l) \th(l-l_n)},
\end{equation}
and its derivative, the differential spectral density, is given by
\begin{equation}
  \r'_l(\l) = {d\r_l(\l)\over dl}
  = {1\over V}\, \vev{\sum_n \d(\l_n-\l)\d(l_n-l)}.
\label{rstr}
\end{equation}
We also introduce a probability
distribution $\cp_\l(l)$ for the localization length $l$ of
eigenmodes with given eigenvalue $\l$ by writing $\r'_l(\l) =
\r(\l) \cp_\l(l)$. This distribution is normalized because, in
finite volume,
\begin{equation}
  \int_{0}^{L} dl\,\r'_l(\l) = \r(\l)
  \quad\text{and hence}\quad
  \int_{0}^{L} dl\, \cp_\l(l) = 1 ,
\label{Pl}
\end{equation}
where $L$ is the (largest) linear size of the lattice.
Below, this upper limit will be implicit.

[In infinite volume there can be truly extended eigenmodes.
Since we expect the eigenmodes at a given $\l$ to be
either all extended or all localized, we have correspondingly
either $\cp_\l(l)=0$ for all finite $l$,
or $\int_{0}^{\infty} dl\, \cp_\l(l)=1$.]

The decay rate of individual localized eigenmodes can be related to
the decay rate of $\G(t)$ for $m_1\to 0$ \cite{lclz}.
Using Eq.~(\ref{lclx}) in Eq.~(\ref{limm1}) gives
\begin{widetext}
\begin{equation}
  m_1 \G(x,y)
  \sim
  \biggl\langle{\sum_n
    \exp\!\left(-{|x-x_n^0|+|y-x_n^0|\over l_n}\right)\,
    {m_1 \over (\l_n - \l)^2 + m_1^2}\biggr\rangle} .
\end{equation}
Averaging over the position of $x_n^0$ gives
\begin{equation}
  m_1 \G(x,y) \sim {1\over V}\, \biggl\langle{\sum_n
    \exp\!\left(-{|x-y|\over l_n}\right)\,
    {m_1 \over (\l_n - \l)^2 + m_1^2}}\biggl\rangle ,
\label{Tinv}
\end{equation}
\end{widetext}
since the average is dominated by locations $x_n^0$ near the straight line
connecting $x$ and $y$.
Hence
\begin{equation}
  \lim_{m_1 \to 0} m_1 \G(x,y)
  \sim
  \p \r(\l) \int dl\, \cp_\l(l)\,
  \exp\!\left(-{|x-y|\over l}\right),
\label{estlim}
\end{equation}
where we have used Eq.~(\ref{rstr}).
Equation~(\ref{Gt}) then gives
\begin{equation}
  \lim_{m_1 \to 0} m_1 \G(t)
  \sim \p \r(\l) \int dl\, \cp_\l(l)\,
  \exp\!\left(-{t\over l}\right).
\label{estt}
\end{equation}
Thus $\Gamma(t)$ decays exponentially, as a weighted average of exponentials
with all localization lengths.
Motivated by this result and ignoring  power corrections,
we finally define the average localization length as
\begin{equation}
  l_\ell(\l) = {1\over \m(\l)},
\label{ll}
\end{equation}
\begin{figure}[thb]
\begin{center}
\includegraphics*[width=.7\columnwidth]{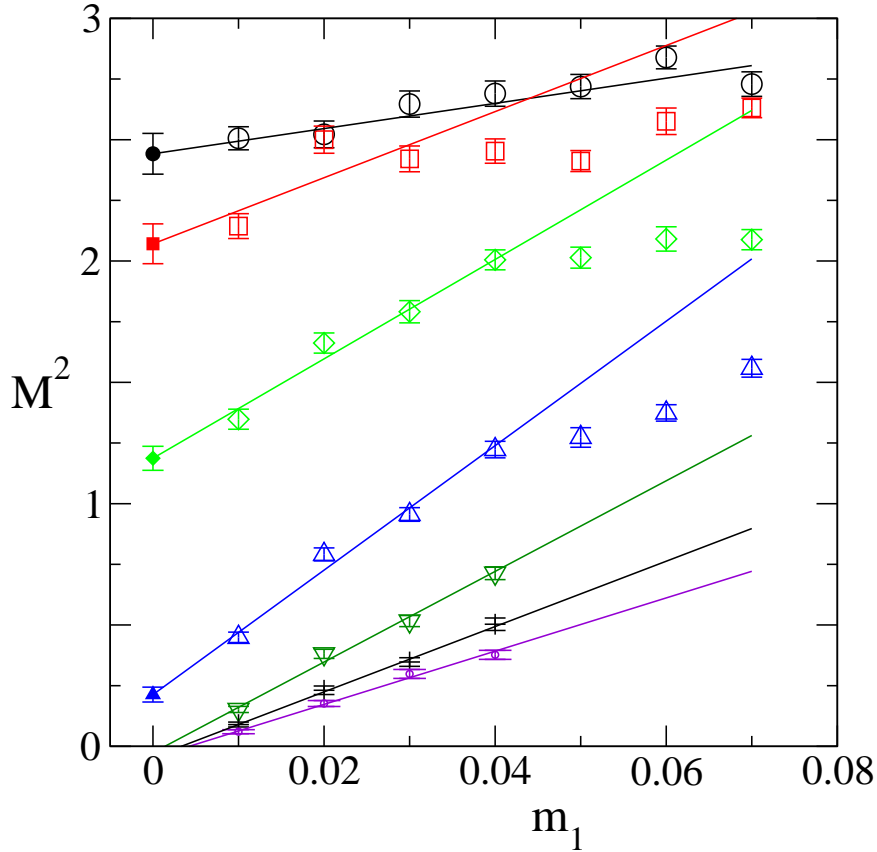}
\caption{Squared decay rate $M^2$ of the two-point function
$\Gamma(t)$ vs.~$m_1$ (open symbols) and its linear
extrapolation to $m_1=0$.
Top to bottom: $\lambda=0$, 0.1, 0.2, 0.3, 0.4, 0.5, 0.6.
The mobility edge is at $\lambda\simeq 0.32$.
All data are for Wilson action at $\beta=5.85$; $m_0=-1.5$.
\label{b5.85_m2_vs_m1}}
\end{center}
\end{figure}

\noindent where $\m(\l)$ is the extrapolation to $m_1=0$
of the decay rate $M$ of the two-point function,
\begin{equation}
  \G(t) \sim \exp[-M(\l,m_1) t], \qquad t \gg a.
\label{mu}
\end{equation}

Equation~(\ref{estt}) actually suggests a slightly different definition of
$\m(\l)$,
based on the extrapolation of the correlation function as a whole,
\begin{equation}
  \lim_{m_1 \to 0} m_1 \G(t) \sim \exp[-\m(\l) t], \qquad t \gg a.
\label{mu'}
\end{equation}
In practice, the extraction of $\m(\l)$ using Eq.~(\ref{mu'}) turns out
to be extremely noisy, but compatible with the results obtained using
Eq.~(\ref{mu}).\footnote{Some further details are provided in Ref.~\onlinecite{mob}.}
We therefore used the definition based on Eq.~(\ref{mu})
throughout our calculations.

We extrapolate to $m_1=0$ by fitting
\begin{equation}
  M^2(\l,m_1) = \mu^2(\lambda) + \alpha(\lambda) m_1.
\label{Mfit}
\end{equation}
See below for an explanation of this choice. We
illustrate this extrapolation in Fig.~\ref{b5.85_m2_vs_m1}. The
tables list the results for $l_\ell=\mu^{-1}$, for comparison with
other characteristic lengths.%
\footnote{We calculated the localization length from
$\Gamma(t)$, without improvement.  Where the tables indicate improvement,
it applies to $\rho$ and $\cal D$ only.}

The derivation leading to Eq.~(\ref{estt}) relies on Eq.~(\ref{limm1}),
where interference effects present in Eq.~(\ref{dblsum}) have been dropped.
This is certainly true when the exponentially localized eigenmodes are
{\em isolated}. A precise definition of what this means
will be given later.
For now it suffices to say that, intuitively,
the eigenmodes with eigenvalue near $\l$ are isolated
if every such eigenmode is supported in a different part of the lattice.

As $\l$ increases, the spectral density grows and so does
the localization length of individual eigenmodes.
Eventually, interference sets in,
meaning that the approximation of Eq.~(\ref{limm1}) is not valid for the $m_1$
values we use. Interference will cause $\G(t)$ to decay faster than
the decay rate of individual eigenmodes.
The definition $l_\ell=\m^{-1}$ will then systematically underestimate
the true (average) localization length, as extracted perhaps by
calculating individual eigenmodes and matching them to Eq.~(\ref{lclx}).
If we are to use Eq.~(\ref{mu}) to calculate the localization length,
we must verify that the eigenmodes
are indeed isolated. We return to this issue below.

\subsection{The mobility edge}
\label{WilB}

As we increase $\l$, we reach a critical value
$\l_c$ where $\m(\l_c)=0$ (see Fig.~\ref{b5.85_m2_vs_m1}).
What is the physical significance of $\l_c$?
According to the preceding discussion, $\m^{-1}(\l)$ provides a reasonable
average value for the localization lengths of individual eigenmodes
if there is no interference, while it underestimates the average
when interference sets in. Either way, $\m(\l)=0$ implies that
the average localization length of individual eigenmodes is infinite;
the eigenmodes have become {\it extended}.
The point $\l_c$ therefore marks the {\it mobility edge}.

Would a different procedure yield a different value for the mobility edge?
One might consider, for instance, an alternative determination based on the numerical calculation of individual eigenmodes.
We believe that such ambiguity can only be an artifact of finite-size effects.
Such effects are perforce significant when $\m(\l) \le 1/L$, since the notion of extended eigenmodes is truly meaningful only in infinite volume.
Thus some disagreement between different determinations
of the mobility edge is to be expected in any finite volume.
Nonetheless, we expect that all methods will converge to the same value in the infinite-volume limit \cite{DJT}.
We are only interested here in obtaining an overall picture, which does not justify the resources that would be needed for, say, a finite-size scaling analysis.

\begin{figure}[htb]
\begin{center}
\includegraphics*[width=.7\columnwidth]{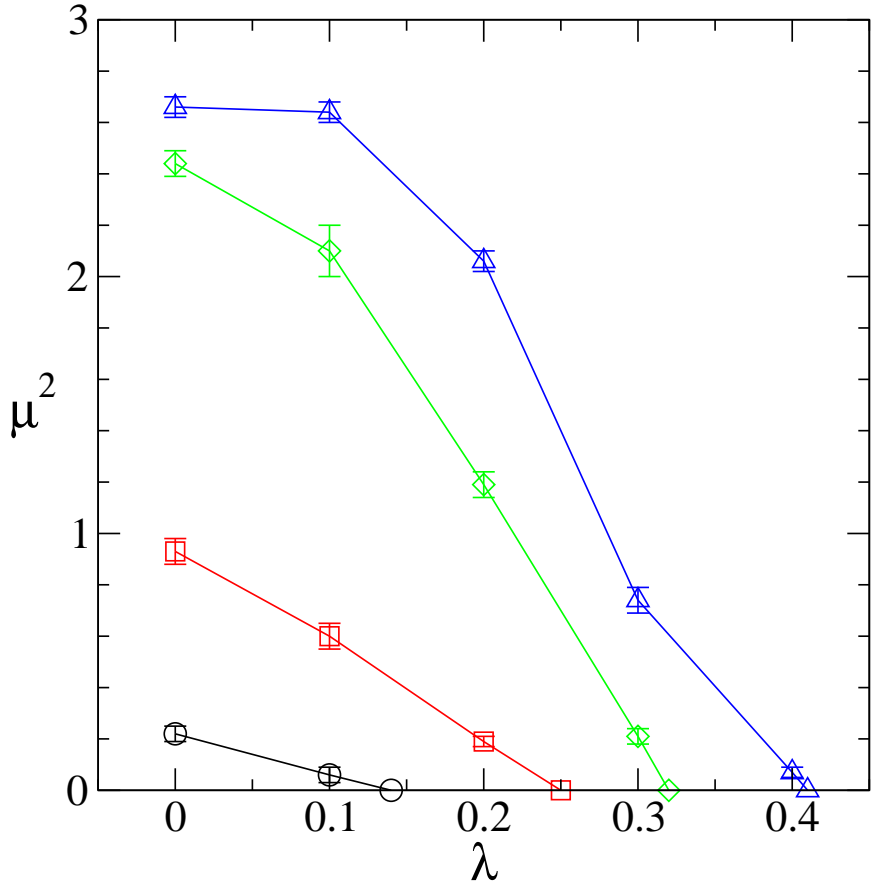}
\caption{The inverse squared localization length $\mu^2$ vs.~$\lambda$.
Top to bottom: $\beta=6.0$, 5.85, 5.7, 5.6.
Intercepts at the $\lambda$ axis are extrapolated from the last two
points in each set, and provide estimates for the mobility edge
$\lambda_c$.
Data are for Wilson action; $m_0=-1.5$.
\label{m2_vs_l_all_beta}}
\end{center}
\end{figure}

\begin{figure}[htb]
\begin{center}
\includegraphics*[width=.7\columnwidth]{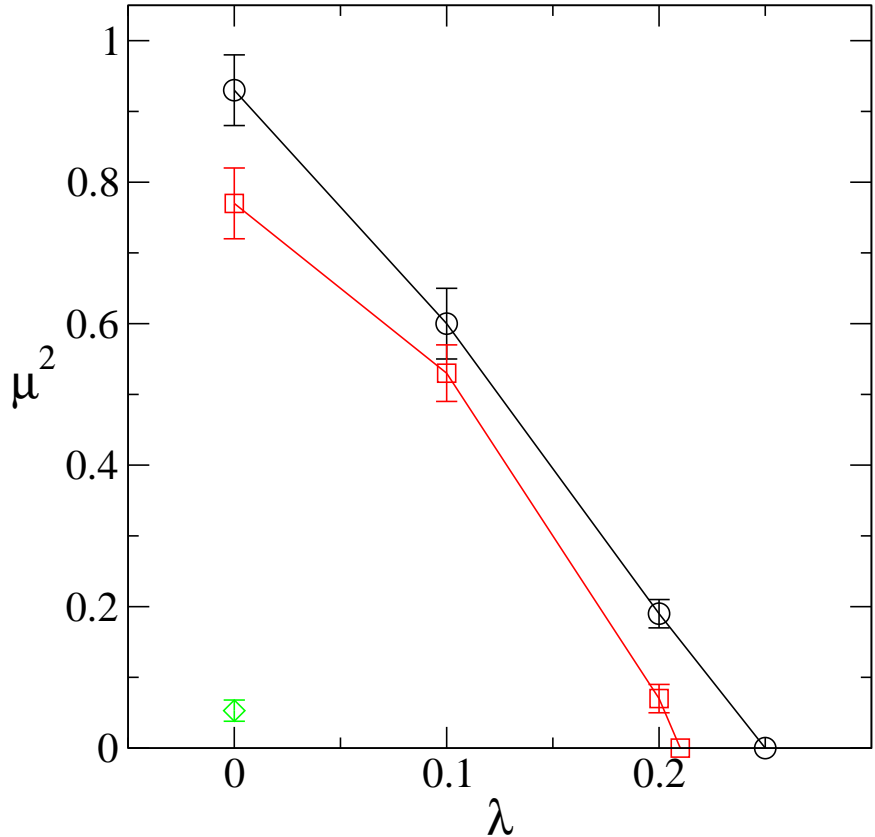}
\caption{Same as Fig.~\ref{m2_vs_l_all_beta}, at $\beta=5.7$. Top
to bottom: $m_0=-1.5$, $-2.0$, $-2.4$.  For $m_0=-2.4$ the
mobility edge is probably very close to $\lambda=0$.  Wilson
action. \label{m2_vs_l_all_kappa}}
\end{center}
\end{figure}

We estimate $\l_c$ simply by a linear extrapolation of $\m(\l)$ to zero, using the last two non-zero
values we have found. The justification for this procedure is
that, for $\l$ near $\l_c$, we expect that $\m(\l)$ will exhibit
the characteristic behavior of a (continuous) phase transition.
See Figs.~\ref{m2_vs_l_all_beta} and~\ref{m2_vs_l_all_kappa}.

Above the mobility edge we expect the $1/m_1$ divergence of $\G(x,y)$
to disappear in the infinite-volume limit; hence Eq.~(\ref{false}) is valid,
and $\tG_\m(p)$ should have a Goldstone pole. We expect that the
physics of these Goldstone poles
to be governed by some effective chiral lagrangian.
Borrowing from the familiar physics of ordinary Goldstone bosons,
we expect $M^2 \propto m_1$ in the presence of the symmetry-breaking
field $m_1$. This is the motivation for the term linear
in $m_1$ in the fit function~(\ref{Mfit}). Below the mobility edge
$M^2$ does not vanish for $m_1\to 0$,
and therefore it is immaterial whether we extrapolate $M$ or $M^2$ to $m_1=0$.
Put together, these considerations suggest that Eq.~(\ref{Mfit})
is an appropriate form both below and above the mobility edge.

Above the mobility edge, the linear extrapolation often gives
a small negative result for
$\mu^2$.
We have neglected to include possible logarithms
which, in the context of a chiral lagrangian, would occur at next-to-leading
order in $m_1$.
These logarithms, together with finite-size corrections, should
move the extrapolated value to zero.

Our linear extrapolation to find the zero of $\mu(\lambda) = l_\ell^{-1}(\l)$
suffers from
an uncertainty that stems from the interference to which we alluded above.
On further analysis (see Sec.~\ref{WilE} below) we will conclude that,
for all $(\beta,m_0)$ where $\lambda_c>0$,
our estimate of $l_\ell$ is unreliable for the largest value of $\lambda$ that
is below the mobility edge $\lambda_c$. (This uncertainly
does not occur for lower values of $\lambda$.)
In fact, the calculated $l_\ell^{-1}$ at this value of $\lambda$ is
likely to be too high; its true value could even be zero.
In other words, this value of $\lambda$ could lie
{\em above\/} the mobility edge.
This means that the last measured point on each curve in
Figs.~\ref{m2_vs_l_all_beta} and~\ref{m2_vs_l_all_kappa} is unreliable.
\subsection{Entering the Aoki phase}
\label{WilC}

Our estimates for the mobility edge $\lambda_c(\beta,m_0)$ are collected in
Table~\ref{table:mobilityedge}.
Consider first the $m_0=-1.5$ results. For reference, we include
the free-theory limit ($\beta=\infty$) \cite{lclz}, where
$\lambda_c$ coincides with the gap of the free $H_W$. At
$\beta=6.0$, $\lambda_c$ is still close to its free-field value.
As we move towards strong coupling, the curve $\lambda_c(\beta)$
steepens before reaching zero somewhere between $\beta=5.6$ and
$\beta=5.5$.
As discussed in Sec.~\ref{Gold}, where $\lambda_c=0$ the two-point function
$\Gamma(t)$ has a zero-momentum pole even at $\lambda=0$.  The
Goldstone Theorem is valid, the Ward identity~(\ref{false}) is
satisfied, and the (quenched) Wilson-fermion theory possesses a Goldstone boson
associated with the spontaneous breakdown of its SU(2) flavor
symmetry. This is the Aoki phase, and we put its boundary between
$\beta=5.6$ and $\beta=5.5$.

Table~~\ref{table:mobilityedge} also shows results at the other
two $m_0$ values for $\beta=5.7$. We find only a small
change between $m_0=-1.5$ and $m_0=-2.0$, while the $m_0=-2.4$ result suggests
that here, too, one is near or within the Aoki phase.

The implications of these results for DWF and overlap fermions
are discussed in Sec.~\ref{DWF}.

\begin{table}[ht]
\caption{ Mobility edge $\lambda_c(\beta,m_0)$. Where no error is
shown, the statistical error is less than one in the last digit.
\label{table:mobilityedge}}
\begin{ruledtabular}
\begin{tabular}{ddd}
\beta & m_0  & \lambda_c \\
\hline
\infty    &  -1.5  &  1/2       \\
6.0       &  -1.5  &  0.41      \\
5.85      &  -1.5  &  0.32      \\
5.7       &  -1.5  &  0.25      \\
5.6       &  -1.5  &  0.14(2)   \\
5.5       &  -1.5  &  0.0       \\
5.4       &  -1.5  &  0.0       \\
\hline
5.7       &  -2.0  &  0.21      \\
5.7       &  -2.4  &  \approx0   \\
\end{tabular}
\end{ruledtabular}
\end{table}

\subsection{Participation number and the support length}
\label{WilD}
Now we turn to a more detailed description of the localized eigenmodes,
for which $|\l|<\l_c$. We begin by defining a measure of the
size of the support of a localized eigenmode.
In the sequel we use this measure to sharpen the notion of isolated localized eigenmodes, and verify the consistency of our approach.

We define \cite{DJT,JLSS} the {\em participation number} $P$ of a normalized
eigenmode $\J(x)$ in $d$ dimensions by
\begin{equation}
  P^{-1} = a^d \sum_x (|\J(x)|^2)^2.
\label{pn}
\end{equation}
The physical meaning is easy to see. Suppose that
the magnitude of the eigenmode density is $|\J(x)|^2 \approx
1/l_s^d$ over a region whose linear size is $l_s$. Then $P^{1/d}
\approx l_s$ is a measure of the linear size of the support of
$\J(x)$. We similarly define a
generalized participation number by
\begin{equation}
  P^{-1}_{d,d'} = a^{d+d'} \sum_{x_{d'+1},\ldots,x_d}
  \Bigl( \sum_{x_1,\ldots,x_{d'}} |\J(x_\m)|^2 \Bigr)^2 .
\label{genPN}
\end{equation}
[Eq.~(\ref{pn}) corresponds to $d'=0$.] In the case
considered we would have
\begin{equation}
  P^{-1}_{d,d'} \approx l_s^{d-d'} ( l_s^{d'-d} )^2 = l_s^{d'-d} .
\label{ldds}
\end{equation}
In the special case $d'=d-1$, we have $P_{d,d-1}
\approx l_s$.

We are thus motivated to define
$l_{s}(n)\equiv P_{4,3}(n)$ to
be the linear size of the support of the $n^{\rm th}$
eigenmode. Substituting in Eq.~(\ref{limm1}) and using translation
invariance together with Eqs.~(\ref{Gt}) and~(\ref{ldds}) we have
\begin{equation}
  m_1 \G(t=0) =
  {1\over V} \biggl\langle{\sum_n
    {1\over l_{s}(n)}\,
    {m_1 \over (\l_n - \l)^2 + m_1^2}}\biggr\rangle+O(m_1).
\label{TinvP}
\end{equation}
Defining
\begin{equation}
  \cd(\l)=\p^{-1}\lim_{m_1 \to 0} m_1 \G(t=0),
\label{defD}
\end{equation}
we have
\begin{equation}
  \cd(\l) = \r(\l) \int dl\, \cp^s_\l(l)\, {1\over l}\,,
\label{lbar'}
\end{equation}
where $\cp^s_\l(l)$ is the probability
distribution for $l_{s}(n)$.
On the basis of
Eq.~(\ref{lbar'}) we define the average {\em support length\/} as
\begin{equation}
  l_s(\l) = {\r(\l) \over \cd(\l)} \,.
\label{ls}
\end{equation}
While we expect the average
localization length $l_\ell$ and the average support length $l_s$
to be quantities of similar magnitude, there is no reason why they
should be the same.

We may also obtain $\cd(\l)$ using the improved estimator~(\ref{impGt}).
Again, results obtained using $\G(t)$ were extrapolated to zero linearly in $m_1$, while those obtained using $\Gimp(t)$ were extrapolated
linearly in $m_1^3$.

We list our results for $l_s$ in the tables. Observe that
below the mobility edge $l_s$ turns out to be always larger than $l_\ell$.
The physical significance of these results is discussed below.

\subsection{Separation distance of localized eigenmodes}
\label{WilE}

We now return to the notion of {\it isolated} localized eigenmodes
mentioned above. The use of Eq.~(\ref{limm1}) depends on neglecting interference among the modes appearing in Eq.~(\ref{dblsum}), which is only justified when $m_1$ is sufficiently small.
We must check whether our smallest values of $m_1$ are indeed sufficiently small.

Spectral sums as in Eq.~(\ref{limm1}) show that $m_1$ is the resolution
with which we detect eigenmodes near $\lambda$.
Let us compare the average support and localization lengths
$l_s$ and $l_\ell$ to the
mean distance between eigenmodes detected at this resolution.
Our smallest value
of $m_1$ is 0.01. The number of eigenmodes with eigenvalues
near $\lambda$ that we detect for a typical gauge
configuration is thus $N(\lambda) \simeq 0.01 V \rho(\lambda)$.
Hence we define
\begin{equation}
  R(\lambda)\equiv[V/N(\l)]^{1/4}\simeq[0.01 \,\rho(\lambda)]^{-1/4}
\label{Rl}
\end{equation}
as a measure of the average separation between eigenmodes
detected at this resolution.
Results for $R(\l)$ are shown in the tables. (We do not quote errors but they can be easily worked out from the $\r(\l)$ data.)
If $l_s \ll R$ the eigenmodes are isolated, and
correlation functions reflect properties of individual localized
eigenmodes, with no interference.%
\footnote{Here it is significant that $l_\ell$ turns out to be smaller than $l_s$.}
 Thus, our method for the extraction of the
average localization and support lengths is valid.

We can also directly estimate the overlap among any two eigenmodes to see whether they interfere.
This overlap satisfies the inequality
\begin{equation}
  |\sbraket{\J_1}{\J_2}|^2 \le \Bigl( a^4\sum_x |\J_1(x)| |\J_2(x)| \Bigr)^2 .
\label{interf}
\end{equation}
We assume $\l_1 \approx \l_2 \approx \l$.
The ``edge'' of the support of $\J_2$ closest to the center of $\J_1$
is at an average distance of around $R-l_s/2$. Using Eq.~(\ref{lclx}) this gives
\begin{equation}
|\sbraket{\J_1}{\J_2}|^2 \alt \exp[-(R-l_s/2)/l_\ell].
\end{equation}
A review of the tables shows the following.\footnote{
  This applies also to results obtained with improved gauge actions,
to be discussed in Sec.~\ref{Imp} below. We ignore those cases where
data for $l_s$ are not available.}
  Leaving out the last value of $\l$ just below the mobility edge we find that
for all other $\l<\l_c$ one has $l_s < R/2$ and $l_\ell < R/5$. Hence
\begin{figure}[thb]
\begin{center}
\includegraphics*[width=.7\columnwidth]{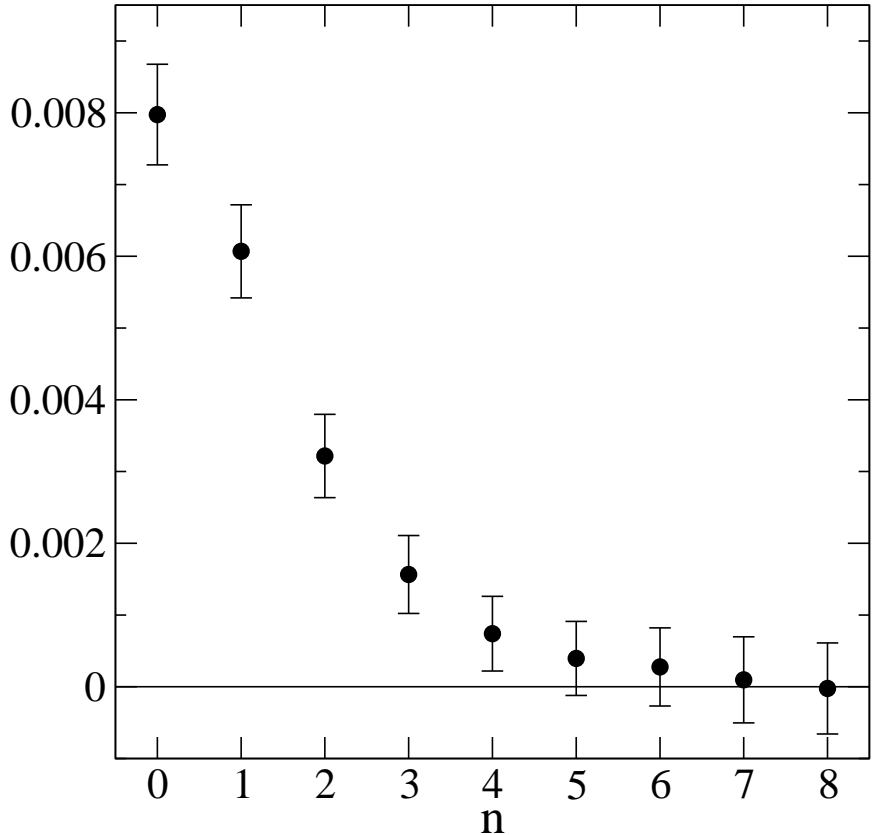}
\caption{Coefficient of the $1/m_1$-divergence in $\tG(\omega_n)$
for $\beta=5.7$, $m_0=-1.5$, $\lambda=0.0$.
\label{GammaT_vs_omega}}
\end{center}
\end{figure}
$|\sbraket{\J_1}{\J_2}|^2 \alt \exp[-(3R/4)/(R/5)]
= \exp(-15/4) \simeq 0.024$.
This demonstrates that, except for the last value of $\l$
just below $\l_c$, all the quantities we measured indeed reflect
properties of individual localized eigenmodes,
with no interference.

Above $\lambda_c$, as well as (in most cases) for the $\l$ value
just below $\l_c$, we have $R \approx l_s \approx l_\ell$.
This means that interference is not a negligible effect,
and the values of $l_s$ and $l_\ell$ no longer
reflect the properties of individual eigenmodes.

\subsection{The $1/m_1$ divergence of the two-point function: $p\ne 0$}
\label{WilF}

Finally, we return to the Ward identity Eq.~(\ref{ppft}). As can be
seen from Eq.~(\ref{noGB}), the localization alternative to Goldstone's
theorem requires that the $1/m_1$ divergence persist for a range
of momenta, and that its coefficient depend smoothly on $p$.

In order to confirm this, we calculated the Fourier transform of $\Gamma(t)$,
\begin{equation}
  \tG(\omega_n) = \sum_t \cos(\omega_n t)
\Gamma(t)
\label{FourierG}
\end{equation}
where $\omega_n=2\pi n/16$, and
extrapolated $m_1 \tG(\omega_n)$ linearly to $m_1=0$. [The
justification for the linear extrapolation is the same as for Eq.~(\ref{rfit}).
We included data for $m_1=0.005$ in the fit.]
The calculation, which provides a consistency check, was done for $\b=5.7$,
$m_0=-1.5$, $\lambda=0$, where the higher spectral density (compared
to $\b=6.0$) makes it easier to obtain good statistics;
the mobility edge here is still far from zero.
We plot the results in Fig.~\ref{GammaT_vs_omega}.
The $\omega$-dependence of the $1/m_1$ divergence is indeed smooth,
as indicated in Eq.~(\ref{noGB}) for small $\omega$.

For comparison, we repeated the calculation for $\lambda=0.5$,
which lies above $\lambda_c$. The extrapolation of
$m_1\tG(\omega_n=0)$ to $m_1=0$ is again straightforward, as it
must be since this gives $\rho(\lambda)$ according to
Eq.~(\ref{sumGt}). Doing the same with $\tG(\omega_n\neq0)$, however,
leads to a huge $\chi^2$, showing that Eq.~(\ref{noGB}) is inapplicable.
This highlights the qualitative difference
between $|\lambda|<\lambda_c$ and $|\lambda|>\lambda_c$.

\section{Improved gauge actions}
\label{Imp}

As an alternative to the Wilson plaquette action, we also studied pure-gauge ensembles generated by the
Iwasaki \cite{iwa} and DBW2 \cite{dbw2} actions,
which have been used in DWF \cite{iwCPPACS,dbw2RBC,ddwf} and overlap
\cite{1gev,gall} simulations.
We set the Wilson mass to $m_0=-1.8$ (see Sec.~\ref{DWF} for explanation of
this choice). For each action, we studied two
values of the lattice spacing, $a^{-1}\simeq 2$~GeV and $a^{-1}\simeq 1$~GeV.
The corresponding bare couplings are $\b=2.6$ and~2.2 respectively for
the Iwasaki action \cite{iwaval}. For the DBW2 action they are
$\b=1.04$ \cite{dbw2RBC} and~0.79 respectively; the latter was determined by interpolation
using $\b$-values from Ref.~\onlinecite{dbw2val}.

\begin{table*}[th]
\caption{%
Spectral properties for Iwasaki gauge action, $\beta=2.6$, $m_0=-1.8$.
The mobility edge is at $\lambda_c\simeq0.38$. Number of configurations
ranges from 600 at $\lambda=0.$ to 100 at $\lambda=0.5$.
\label{tab2.6}}
\begin{ruledtabular}
\begin{tabular}{dddddd}
\lambda & \rho & l_\ell & \cd & l_s & R  \\
\hline\hline
0.0\footnotemark[1] & 7.(3)\times 10^{-7}  & 0.72(1) & -                    & -     & 110.  \\
0.1\footnotemark[1] & 1.8(5)\times 10^{-5} & 0.73(1) & 4.(1)\times 10^{-6}  & 4.(2) & 50.   \\
0.2\footnotemark[1] & 2.3(9)\times 10^{-4} & 0.80(1) & 1.3(7)\times 10^{-4} & 2.(1) & 25.   \\
0.3\footnotemark[1] & 9.(2)\times 10^{-4}  & 1.20(2) & 2.7(8)\times 10^{-4} & 3.(1) & 18.   \\
\hline
0.4\footnotemark[1] & 0.111(2)             & \infty  & -                    & -     & 5.5   \\
0.5\footnotemark[1] & 0.277(5)             & \infty  & 0.040(4)             & 7.(7) & 4.3   \\
\end{tabular}
\end{ruledtabular}
\footnotetext[1]{Results from improved estimator.}
\end{table*}
\begin{table*}[th]
\caption{%
Spectral properties for Iwasaki gauge action, $\beta=2.2$, $m_0=-1.8$.
The mobility edge is at $\lambda_c\simeq0.27$. Measurements on
200 gauge configurations.
\label{tab2.2}}
\begin{ruledtabular}
\begin{tabular}{dddddd}
\lambda & \rho & l_\ell & \cd & l_s & R  \\
\hline\hline
0.0\footnotemark[1] &  0.0063(2)  &  0.80(1)  &  0.0021(1)  &  3.0(2) &  11.    \\
0.1\footnotemark[1] &  0.0095(3)  &  0.91(2)  &  0.0032(2)  &  3.0(2) &  10.    \\
0.2 &  0.034(2)   &  1.40(5)  &  0.010(2)   &  3.4(7) &   7.4    \\
\hline
0.3 &  0.120(5)   &  \infty   &  0.025(4)   &  5.(1)  &   5.4   \\
0.4 &  0.27(1)    &  \infty   &  0.017(7)   & 16.(10) &   4.4   \\
\end{tabular}
\end{ruledtabular}
\footnotetext[1]{Results from improved estimator.}
\end{table*}

\clearpage
\pagebreak

\begin{figure}[htb]
\begin{center}
\includegraphics*[width=.7\columnwidth]{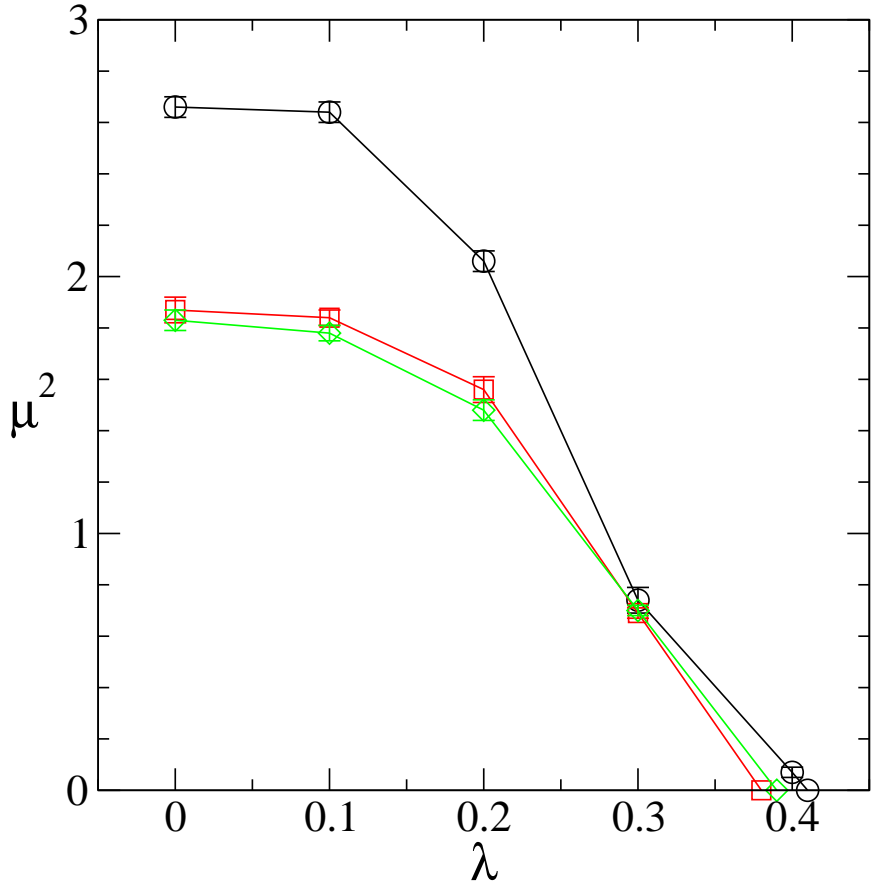}
\caption{The inverse squared localization length $\mu^2$ vs.~$\lambda$,
as in Fig.~\ref{m2_vs_l_all_beta}.
Comparison of three gauge actions at $a^{-1}\simeq 2$~GeV:
Wilson action at $\beta=6.0$, with $m_0=-1.5$ (circles);
Iwasaki action at $\beta=2.6$, with $m_0=-1.8$ (squares);
DBW2 action at $\beta=1.04$, with $m_0=-1.8$ (diamonds).
\label{m2_vs_l_3actions_2GeV}}
\end{center}
\end{figure}

\begin{figure}[htb]
\begin{center}
\includegraphics*[width=.7\columnwidth]{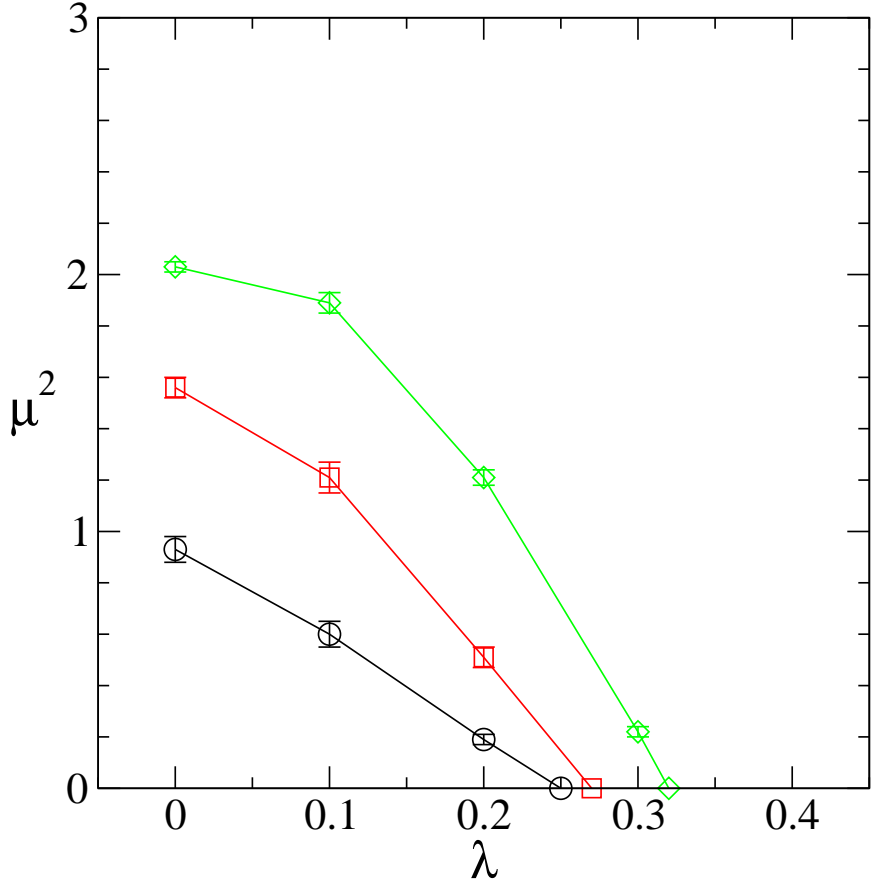}
\caption{As in Fig.~\ref{m2_vs_l_3actions_2GeV}, but at $a^{-1}\simeq 1$~GeV:
Wilson action at $\beta=5.7$, with $m_0=-1.5$ (circles);
Iwasaki action at $\beta=2.2$, with $m_0=-1.8$ (squares);
DBW2 action at $\beta=0.79$, with $m_0=-1.8$ (diamonds).
\label{m2_vs_l_3actions_1GeV}}
\end{center}
\end{figure}

We present our results in Tables 10--13 and Figs.~\ref{m2_vs_l_3actions_2GeV}
and~\ref{m2_vs_l_3actions_1GeV}.
A comparison of the (inverse squared) localization length for the three
gauge actions at 2~GeV is shown in  Fig.~\ref{m2_vs_l_3actions_2GeV}.
The Iwasaki and DBW2 localization lengths are almost the same, an indication of the good scaling properties of the two actions.
The mobility edge is almost the same for all three actions.
Our results at 1~GeV are compared in Fig.~\ref{m2_vs_l_3actions_1GeV}.
There is a gradual decrease in the localization length
from Wilson to Iwasaki to DBW2. Differences in the value of
the mobility edge, while bigger than in the 2~GeV case, remain small.

As can be seen from Tables 10--13,
the main difference among the three actions is a dramatic reduction
in the low-$\l$ spectral density as we go from the Wilson action
to the Iwasaki and then to the DBW2 gauge actions.
What distinguishes the three actions is the coefficient
of the $1\times 2$ rectangle, which is zero for the standard plaquette action
and in a common parametrization \cite{iwCPPACS,dbw2RBC}
is $c=-0.331$ for the Iwasaki action and $c=-1.4069$ for the DBW2 action.
We see that the decrease in the low-$\l$ spectral density is
correlated with a more negative rectangle coefficient $c$.
The rectangle term in the action suppresses the small-size dislocations
\cite{dbw2,dbw2RBC} responsible for the existence of the near-zero modes.

At (or just above) the mobility edge the spectral density remains of the same
order as for the plaquette action: $\r(\l_c) \approx 0.1$.
This means that, unlike the plaquette action,
the Iwasaki and DBW2 spectral densities rise steeply as $\l$
approaches the mobility edge from below. At 2~GeV,
a comparison of the spectral-density values just above and below
the mobility edge shows a jump by two orders of magnitude
for both improved actions.

\begin{table*}[th]
\caption{%
Spectral properties for DBW2 gauge action, $\beta=1.04$, $m_0=-1.8$.
The mobility edge is at $\lambda_c\simeq0.39$. Between 200 and 400
gauge configurations were used.
\label{tab1.04}}
\begin{ruledtabular}
\begin{tabular}{dddddd}
\lambda & \rho & l_\ell & \cd & l_s & R  \\
\hline\hline
0.0\footnotemark[1] & -                    &  0.74(1) & - & - &    -  \\
0.1\footnotemark[1] & 2.(1)\times 10^{-7}  &  0.75(1) & - & - &  150. \\
0.2\footnotemark[1] & 6.(2)\times 10^{-6}  &  0.82(1) & - & - &   65. \\
0.3\footnotemark[1]\footnotemark[2] & 3.4(7)\times 10^{-4} &  1.2(2)  & - & - &   23. \\
\hline
0.4 & 0.058(2)             &  \infty  & - & - &    6. \\
\end{tabular}
\end{ruledtabular}
\footnotetext[1]{Results from improved estimator.}
\footnotetext[2]{Extrapolation with Eq.~(\ref{cubfit}) failed.
We added terms linear and quadratic in $m_1$.}
\end{table*}
\begin{table*}[th]
\caption{%
Spectral properties for DBW2 gauge action, $\beta=0.79$, $m_0=-1.8$.
The mobility edge is at $\lambda_c\simeq0.32$. Measurements on
200 gauge configurations.
\label{tab0.79}}
\begin{ruledtabular}
\begin{tabular}{dddddd}
\lambda & \rho & l_\ell & \cd & l_s & R  \\
\hline\hline
0.0\footnotemark[1] &  0.0013(1)  &  0.70(1)  &  0.00041(5)  &  3.2(5) &  17.   \\
0.1\footnotemark[1] &  0.0023(2)  &  0.73(1)  &  0.0009(1)   &  2.6(4) &  14.   \\
0.2\footnotemark[1]\footnotemark[2] &  0.0064(4)  &  0.91(1)  &  0.0024(2)   &  2.7(3) &  11.   \\
0.3\footnotemark[1] &  0.058(1)   &  2.1(1)   &  0.012(1)    &  4.8(4) &   6.4  \\
\hline
0.4 &  0.22(1)    &  \infty   &  0.030(5)    &  7.(1)  &   4.6  \\
\end{tabular}
\end{ruledtabular}
\footnotetext[1]{Results from improved estimator.}
\footnotetext[2]{Extrapolation with Eq.~(\ref{cubfit}) failed.
We added an $m_1^2$ term.}
\end{table*}


\section{Discussion}
\label{DWF}

Our numerical results bear, first, on the phase structure of the
super-critical Wilson operator in various ensembles; and, second,
on the overlap and DWF theories based on the Wilson operator.
We examine each of these in turn.

\subsection{The Aoki phase diagram}

The Aoki phase diagram has dictated our list of $(\b,m_0)$ values
for calculation, and our results in turn yield information about
the diagram.
We can now discuss in more detail the choice of $(\b,m_0)$ values
we have made.

There is considerable evidence (see \eg Ref.~\onlinecite{SCRI}) that,
for plaquette actions with $\b=5.7$ to~6.0,
the low-$\l$ spectral density of the Wilson operator varies slowly
over the range $-2.0 < m_0 < -1.5$. In view of this,
we have chosen to carry out most
of our calculations with the plaquette action for the single value $m_0=-1.5$.
We have tested this insensitivity to $m_0$ by considering as well
$\b=5.7$, $m_0=-2.0$.
We see in Fig.~\ref{m2_vs_l_all_kappa} that the localization lengths~$l_\ell$ for all $\l$
are almost unchanged as $m_0$ changes from~$-1.5$ to~$-2.0$; hence, the mobility edge is unchanged as well.
The rough equality
extends to the support length and to the spectral density for
$\l$ sufficiently below the mobility edge, which means in practice $\l\le 0.1$
(see Tables~\ref{tab5.7} and~\ref{tab5.7M2.0} as well as
Fig.~\ref{rho_vs_l_all_kappa_linear}).

Old results \cite{aokiq} on the Aoki phase diagram indicate that the point
$(\b=6.0,m_0=-2.4)$ lies in the second ``finger'' of the Aoki phase.
Our results show that the point $(5.7,-2.4)$ is indeed at a boundary
of this finger or inside it.

Looking ahead to DWF,
our results indicate that, for $\beta\agt6.0$, one can use any value
of $m_0$ in the range $[-2.0,-1.5]$.
(The same applies to the overlap construction.)
Mean-field arguments suggest that the optimal $m_0$
(related to the ``domain-wall height'' $M$ via $M=-m_0$) for DWF simulations
is roughly $m_0 \approx -1 + m_c(g_0)$ \cite{DWFheight}.
The commonly used value at $a^{-1} \simeq 2$~GeV
is $M = -m_0 = 1.8$. This is consistent with the mean-field estimate and lies
comfortably in the middle of the range of $m_0$ we considered.
In view of our evidence for the insensitivity of the key spectral properties
over this range, we have chosen the value
$m_0 = -1.8$ also for the improved-action calculations.

Our scan of $\beta$ values for $m_0=-1.5$ places the boundary of the Aoki phase of the plaquette action between $\beta=5.5$ and $\beta=5.6$ (cf.~\cite{aokiq}).

\subsection{Implications for overlap fermions}

The overlap operator is constructed explicitly in terms of
the Wilson operator $H_W$,
\begin{equation}
\dov=1-\gamma_5{H_W\over |H_W|}.
\label{dov}
\end{equation}
Our knowledge of the spectrum of $H_W$
is thus of immediate import to overlap
calculations; in particular,
it is critical to understanding the locality of the operator.
The long-distance tail of $\dov$ has two components \cite{HJL,lclz}, and both must be kept under
control.
One component comes from the near-zero modes; its range is
the localization length $l_\ell(\bar\lambda)$, where $\bar\lambda$ will be defined shortly.
The other component comes from modes near the mobility edge, and its range
is the inverse of $\lambda_c$.

We denote by $\lov$ the range of the overlap operator:
For $|x-y|\gg a$ one has $|\dov(x,y)| \sim \exp(-|x-y|/\lov)$.
In Ref.~\onlinecite{lclz} we discussed the different roles of the extended and
the localized modes in determining $\lov$,
and arrived at an estimate for the contribution of the localized spectrum.%
\footnote{This estimate complements the discussion of Ref.~\onlinecite{HJL}.
The analytical techniques of Ref.~\onlinecite{HJL} are inapplicable to the
localized spectrum because it becomes dense in the infinite-volume limit.}
Here we revisit the issue in the light of our numerical findings.

In the treatment of the localized eigenmodes
the essential assumption was the absence of interference between
different modes. This should be true, in particular, for the low-lying
modes that make up the \textit{mutually isolated} subspace.
These are
all the localized eigenmodes in the interval  $[-\bar\l,\bar\l]$
where $0 < \bar\l < \l_c$ is chosen such
that for any two eigenvalues $\l_m,\l_n \in [-\bar\l,\bar\l]$,
the corresponding localized eigenmodes are isolated (on average) in the sense
of Sec.~\ref{WilE}.
The contribution of the mutually isolated subspace to the
ensemble-averaged overlap operator
is then estimated to be%
\footnote{ The introduction of $\bar\l$ improves on the discussion of
Ref.~\onlinecite{lclz}, where the rather arbitrary cutoff $\l_c/2$
was imposed on the integral [see in particular Eq.~(6.3) therein].}
\begin{equation}
  \svev{|\dov(x,y)|}_{|\l|\le \bar\l}
  \approx \int_{-\bar\l}^{\bar\l} d\l\, \r(\l)
  \exp\left(-{|x-y|\over 2l_\ell(\l)}\right).
\label{iso}
\end{equation}
Since both $\r(\l)$ and $l_\ell(\l)$ are monotonically increasing,
the integral is dominated by its upper limit, \textit{viz.}
\begin{equation}
  \svev{|\dov(x,y)|}_{|\l|\le \bar\l}
  \approx \bar\l \r(\bar\l)
  \exp\left(-{|x-y|\over 2l_\ell(\bar\l)}\right).
\label{iso'}
\end{equation}
This is the first of the two pieces determining $\lov$.

We may arrive at a numerical estimate as follows.
Consider for definiteness our results for the Wilson gauge action
at $\b=6.0$, shown in Table~1.
A conservative guess for the edge of the mutually isolated spectrum is
$\bar\l \approx 0.2$.
Since $\r(0.2)$ is of the order of $10^{-3}$, we find that
the integrated spectral density $\int_{-\bar\l}^{\bar\l} d\l\, \r(\l)$
is of the order of $10^{-4}$.
The total number of eigenvalues in the interval  $[-\bar\l,\bar\l]$
is given by the volume $V$ times this integral,
and so the mean space-time separation $\bar{R}$ between
any two eigenmodes in this spectral interval%
\footnote{
Note that $\bar R$ measures the average separation between
{\it all} modes in the interval $[-\bar\lambda,\bar\lambda]$,
whereas $R(\lambda)$ defined in Sec.~\ref{WilE} measures the average
separation between modes in an $O(m_1)$ neighborhood around $\lambda$.}
 is $\bar{R} \simeq 10$.
Using similar considerations to Sec.~\ref{WilE},
this corresponds to $|\sbraket{\J_n}{\J_m}|^2=O(0.01)$,
which is certainly a small overlap.
All the modes in this spectral interval are, therefore, mutually isolated.

Evaluating Eq.~(\ref{iso'}) for $\bar\l=0.2$ gives
\begin{equation}
  \svev{|\dov(x,y)|}_{|\l|\le \bar\l}
  \approx 10^{-4} \exp\left(-{|x-y|\over 1.4}\right).
\label{iso0.2}
\end{equation}
This semi-quantitative estimate highlights two points concerning $\bar\l$.
First, the small prefactor signifies a large value of $\bar R$.  The latter must be appreciably larger than the support length $l_s$ in order for the modes to be mutually isolated, and $\bar\l$ must be chosen to make it so.
Moreover, pushing $\bar\l$
to a somewhat higher value would also increase $l_\ell(\bar\l)$
and thus modify the coefficient in the exponent.
Equation~(\ref{iso0.2}) still provides a rough idea of
the contribution of the mutually isolated localized modes.

The modes with $|\l_n|\ge \bar\l$ are of two types:
the localized eigenmodes
in the interval $\bar\l \le |\l_n| \le \l_c$, and the extended modes
above $\l_c$. The rise in the spectral density near
and above $\l_c$ leads us to expect that the resulting contribution
to $\lov$ will be controlled by $\l_c$ itself.
For these modes the analysis of Ref.~\onlinecite{HJL} applies, and
a comparison of our results to theirs supports this conjecture.
For $\b=6.0$ and $m_0=-1-s=-1.4$ and $-1.6$, Table~1 in Ref.~\onlinecite{HJL}
reports $\lov^{-1}\simeq 0.49$
and~0.45, respectively.
This can be compared to our result for the mobility edge at $m_0=-1.5$,
namely,
$\l_c\simeq0.41$.
Given the differences in the details,
the close agreement may be a numerical coincidence.
Still, this suggests that the contribution of the rest of the spectrum
may indeed be dominated by the extended modes just above the mobility edge,
and that
\begin{equation}
  \svev{|\dov(x,y)|}_{|\l|\ge \bar\l}
  \approx \cc \exp\left(-\l_c |x-y|\right),
\label{rest}
\end{equation}
where $\cc=O(1)$.
This is the second contribution to the tail of $\dov$.

As explained above,
the prefactor in Eq.~(\ref{iso0.2}) is bound to be small.
Comparing Eqs.~(\ref{iso0.2}) and~(\ref{rest})
suggests that $\lov$ is dominated
by the mobility edge and not by the near-zero modes.
This is consistent with the results of Ref.~\onlinecite{fat}.
There it is found that changing the gauge action from Wilson to Iwasaki
or to DBW2 at quenched $a^{-1} \simeq 2$~GeV has little effect on
$\lov$. Indeed, while the spectral density
of the localized modes is a sensitive function of the gauge action,
the mobility edge itself varies little among the three gauge actions
(see Tables 10--13 and Fig.~\ref{m2_vs_l_3actions_2GeV}).
Of course, if $2l_\ell(\bar\lambda)>\lambda_c^{-1}$ the small contribution of
the near-zero modes will dominate the asymptotic tail of $\dov$ at
large distances.
For all cases we study here, however, the opposite inequality holds.

The overlap kernel's decay rate $\lov^{-1}$
can be interpreted as the mass of unphysical degrees of freedom.
In practice, these degrees of freedom can be uncomfortably light.
Assuming $\lov \approx \l_c^{-1}$,
our data allow us to estimate their mass for the pure-gauge ensembles.
When the cutoff is $a^{-1} \simeq 2$~GeV, we find
$\lov^{-1} \simeq 0.4 \times 2~{\rm GeV} =800$~MeV
for all three gauge actions.%
\footnote{This rough equality among the actions is in agreement with the above mentioned results
of Ref.~\onlinecite{fat}.}
This is presumably a high enough scale
to qualify as part of the discretization errors.
For $a^{-1} \simeq 1$~GeV, on the other hand, we obtain
$\lov^{-1} \simeq 0.25 \times 1~{\rm GeV} =250$~MeV
for the Wilson action, 270~MeV for Iwasaki, and 320~MeV for DBW2.
This is an alarmingly low scale for unphysical degrees of freedom.%
\footnote{
Note in particular the 1~GeV overlap simulations
of Ref.~\onlinecite{1gev}.
}

\subsection{Implications for domain-wall fermions}

Our discussion of DWF will be less detailed for
two reasons. First, the theoretical background has already been discussed in
detail \cite{lclz,next,mres,mresRBC}.
Also, the relevant ``hamiltonian'' for DWF is not $H_W$ itself
but the logarithm of the fifth-dimension transfer matrix,
a different (though closely related) operator.

DWF achieve exact chirality when the number of sites $L_5$ in the fifth
dimension tends to infinity. DWF simulations are performed
at finite $L_5$, typically in the range of 10--20.
The main question is what is the size of chiral symmetry violations
due to the finiteness of $L_5$.
A quantitative measure of these violations
is provided by the residual mass $\mres$ \cite{iwCPPACS,dbw2RBC,mresRBC},
which is the small additive correction to the quark mass
determined from the PCAC relation~\cite{dwf}.
(Alternatively, $\mres$ can be determined from the extrapolation
of the pion mass to the chiral limit.)
Here, too, there are two terms that can be ascribed to extended and
localized modes,
\begin{equation}
  \mres \approx \cc_1 \exp(-\tilde\l_c L_5) + {\cc_2 \over L_5} ,
\label{mres}
\end{equation}
where $\cc_1=O(1)$ comes from the extended modes near the mobility edge and
\begin{equation}
  \cc_2 \approx L_5 \int_{-1/L_5}^{1/L_5} d\l\, \tilde\r(\l)
  \approx \tilde\r(1/L_5)
\label{mreslcl}
\end{equation}
comes from the low-lying, localized modes.
Because of the rapidly growing spectral density, the localized modes'
contribution is dominated by modes with $|\l_n| \approx 1/L_5$;
we ignore a power-law correction to the
extended modes' contribution.%
\footnote{
We thank N.~Christ for a discussion of Eq.~(\ref{mreslcl}).
This formula, derived from
Appendix~C.2 of Ref.~\onlinecite{next}, provides a slightly better estimate
of the contribution of the localized modes than that given in
Ref.~\onlinecite{mres}.
Equation~(\ref{mres}) corrects the discussion of the
near-zero modes contribution to $\mres$ given in  Ref.~\onlinecite{lclz}.
In particular, Eq.~(6.12) therein as well as the first term on the
right-hand side of Eq.~(7.1) there are erroneous.
The limits of integration in Eq.~(\ref{mreslcl}) here come of demanding that $\exp(-\l L_5)=O(1)$ for modes included in the integral.
}
The tildes indicate spectral quantities of the new ``hamiltonian,''
\begin{equation}
  \tH = -\log(T^2)/(2a_5) ,
\label{tH}
\end{equation}
where $T(a_5)$ is the transfer matrix for hopping in the
fifth direction and $a_5$ is the corresponding lattice spacing
(conventional DWF have $a_5=1$).

The ``hamiltonians'' $H_W$ and $\tH$
share identical zero modes \cite{dwf}.
Thus we expect that for the near-zero modes,
the spectral density (and other properties) of $\tH$ is fairly close to
that of $H_W$. Replacing $\tilde\r(\l)$ by $\r(\l)$
in Eq.~(\ref{mreslcl}) should yield a reasonably good approximation.
On the other hand, further away from $\l=0$ the spectra of $\tH$ and $H_W$
need not be equal. In particular the mobility edges could be quite different.
Nonetheless, again because of the identity of the zero modes,
the mobility edges of $\tH$ and $H_W$ reach zero simultaneously.
Thus the same Aoki phase defines the forbidden region for both overlap and DWF.

As noted, the numerical results of this paper are relevant for
determining the near-zero modes' contribution to $\mres$,
but separate calculations would be needed to determine $\tilde\l_c$,
the mobility edge of $\tH$, which governs the extended modes' contribution
to $\mres$. One can alternatively determine these quantities by
calculating the residual mass as a function of $L_5$
\cite{mresRBC,iwCPPACS,dbw2RBC} and fitting to Eq.~(\ref{mres}).

Equation~(\ref{mres}) shows that, unlike overlap fermions,
the physics of finite-$L_5$ DWF simulation and, in particular,
the value of $\mres$,
are sensitive functions of the near-zero modes of $\tH$ (or of $H_W$).
In quenched DWF simulations, using the Iwasaki or DBW2 actions
allows reaching negligibly small values of $\mres$ \cite{iwCPPACS,dbw2RBC}.
For overlap simulations, on the other hand,
these gauge actions are advantageous
for a technical reason: overlap simulations require an exact treatment
(within numerical precision)
of all the Wilson eigenvalues in a certain interval $[-\d,\d]$,
and reducing the number of modes in this interval speeds up
the simulation. We note that there exist versions of DWF,
in particular the so-called M\"obius fermions \cite{mobius},
where the near-zero modes' contribution to $\mres$ decreases
much faster with $L_5$ (see also~\cite{Deff,AB,JN}).

\subsection{Summary}

The continuum limit of lattice QCD with either
DWF or overlap fermions can be taken while letting
$m_0\to -1$. Since all correlation lengths associated with the Wilson operator
itself remain finite in lattice units, there is little doubt that the
continuum limit is correct, and that no unphysical excitations
can survive it. Issues addressed in this paper have to do with MC simulations
at finite lattice spacing.

Using Green function techniques, we have determined the mobility edge
of the hermitian Wilson operator for a number of pure-gauge ensembles
with plaquette, Iwasaki and DBW2 gauge actions. Our results
allow mapping a portion of the (quenched) Aoki phase diagram.
Where the mobility edge is non-zero, we have also characterized
the localized spectrum in terms of an average support length and an
average localization length, the latter determined from
the asymptotic decay rate of the mode density.

Our results are of direct relevance to the overlap operator.
For the near-zero modes,
or, more precisely, for the mutually-isolated subspace of the localized modes,
we found that the localization length
is consistently smaller than the support length.
We have also found that twice the localization length is not
bigger than the inverse mobility edge. Together, these findings imply
that the near-zero modes
play little role in setting the range of the overlap operator.

We argue that the range of the overlap operator
is set by, and is roughly equal to, the inverse mobility edge.
Our results for the mobility edge suggest that it is fairly safe to perform
(quenched) overlap simulations at $a^{-1}=2$~GeV.
The same is not true when the cutoff is 1~GeV;
for all three gauge actions we find that the standard overlap operator
is likely to contain unphysical quark-like degrees of freedom
as light as 250--300~MeV.
This casts serious doubt on the validity of 1~GeV overlap simulations.
In general, the determination of the range of the overlap operator
is an important test that must be carried out for any new
overlap simulation.

Closely related to the mobility edge is the mass of the lowest pseudoscalar excitation of the Wilson fermion action---Wilson's original would-be pion.  It, too, is a non-physical excitation where the overlap operator is concerned.  Where we demand that couplings be chosen such that the mobility edge is far from zero, the same can be said of the mass of the Wilson pion.%
\footnote{If it turns out that the A and C regions in Fig.~\ref{aoki_crude} are separated by first-order transitions instead of fingers, the Wilson pion may still have a very small mass near the transitions \cite{ShSi,GSS}.}

Obtaining the corresponding information for DWF will require
the study of a different ``hamiltonian,''
the logarithm of the fifth-dimension transfer matrix, $\tilde{H}$.
In particular, it is important to study the range of the effective
four-dimensional operator $D_{\rm eff}$ obtained by integrating out
the five-dimensional bulk modes and the pseudofermions \cite{Deff},
and to determine how the range of $D_{\rm eff}$ is affected by $\tilde{H}$ as a function of $L_5$ and $a_5$.
The near-zero spectrum of $\tilde{H}$ is similar
to that of the Wilson operator, and we are thus able to confirm the picture
that the near-zero modes (of either $H_W$ or $\tilde{H}$) make a major
contribution to the residual mass.

Last, we note that numerical simulations with dynamical DWF \cite{ddwf}
(or dynamical overlap fermions)
require the largest available computers.
Not only production runs, but also exploratory runs are very expensive.
The process of closing in on an optimal set of simulation parameters
can greatly benefit from the (low-cost) determination
of the quantities studied here---the mobility edge and
the characteristics of the localized modes.

\begin{acknowledgments}
Our computer code is based on the public lattice gauge theory code
of the MILC collaboration, available from
http://www.physics.utah.edu/${\scriptstyle\!\sim\,}$detar/milc/.
We thank the Israel Inter-University Computation Center for a
grant of time on supercomputers operated by the High Performance Computing Unit. Additional computations were
performed on a Beowulf cluster at San Francisco State University.
This work was supported by the Israel Science Foundation under
grant no.~222/02-1, the Basic Research Fund of Tel Aviv
University, and the US Department of Energy.
\end{acknowledgments}

\end{document}